\RequirePackage{fix-cm}
\documentclass[pdftex,twocolumn,epjc3,final]{svjour3}

\RequirePackage{graphicx}
\RequirePackage[T1]{fontenc}
\RequirePackage{subfig}
\RequirePackage{mathptmx}
\RequirePackage{helvet}
\RequirePackage{color}
\RequirePackage{url}
\RequirePackage{xspace}
\RequirePackage{booktabs}
\RequirePackage{amsmath}
\RequirePackage{algorithm}
\RequirePackage[noend]{algpseudocode}
\RequirePackage{pifont}

\begin{document}
    \title{The Pandora Software Development Kit for Pattern Recognition}
    \author{J. S. Marshall\thanksref{e1,addr1} \and M. A. Thomson\thanksref{addr1} }
    \thankstext{e1}{e-mail: marshall@hep.phy.cam.ac.uk}
    \institute{Cavendish Laboratory, University of Cambridge, Cambridge, United Kingdom\label{addr1} }
    %\date{Received: date / Accepted: date}
    % The correct dates will be entered by the editor

\maketitle
    \begin{abstract}
        The development of automated solutions to pattern recognition problems is important in many areas of scientific research and human endeavour. This paper describes the
        implementation of the Pandora Software Development Kit, which aids the process of designing, implementing and running pattern recognition algorithms. The Pandora
        Application Programming Interfaces ensure simple specification of the building-blocks defining a pattern recognition problem. The logic required to solve the problem
        is implemented in algorithms. The algorithms request operations to create or modify data structures and the operations are performed by the Pandora framework. This design
        promotes an approach using many decoupled algorithms, each addressing specific topologies. Details of algorithms addressing two pattern recognition problems in High Energy
        Physics are presented: reconstruction of events at a high-energy $e^{+}e^{-}$ linear collider and reconstruction of cosmic ray or neutrino events in a liquid argon
        time projection chamber.
    % \PACS{PACS code1 \and PACS code2 \and more}
    % \subclass{MSC code1 \and MSC code2 \and more}
    \end{abstract}

%%%%%%%%%%%%%%%%%%%%%%%%%%%%%%%%%%%%%%%%%
% Main part
%%%%%%%%%%%%%%%%%%%%%%%%%%%%%%%%%%%%%%%%%

\section{Introduction}
\label{sec::Introduction}

Pattern recognition is the identification of structures or regularities in data. Problems requiring a pattern recognition solution occur in all areas of scientific research and
our everyday lives. This document describes the implementation of the Pandora Software Development Kit (SDK), which aims to ease the process of designing, implementing and
running pattern recognition algorithms. The Pandora SDK was created to address the problem of identifying energy deposits from individual particles in fine granularity 
detectors in High Energy Physics (HEP). The ideas described in this document are, however, actually quite generic, covering a wide array of problems where the aim is to sort
points in time or space into higher-level structures.

\begin{figure}[!h]
  \begin{center}
     \subfloat[][]{\includegraphics[width=0.45\textwidth]{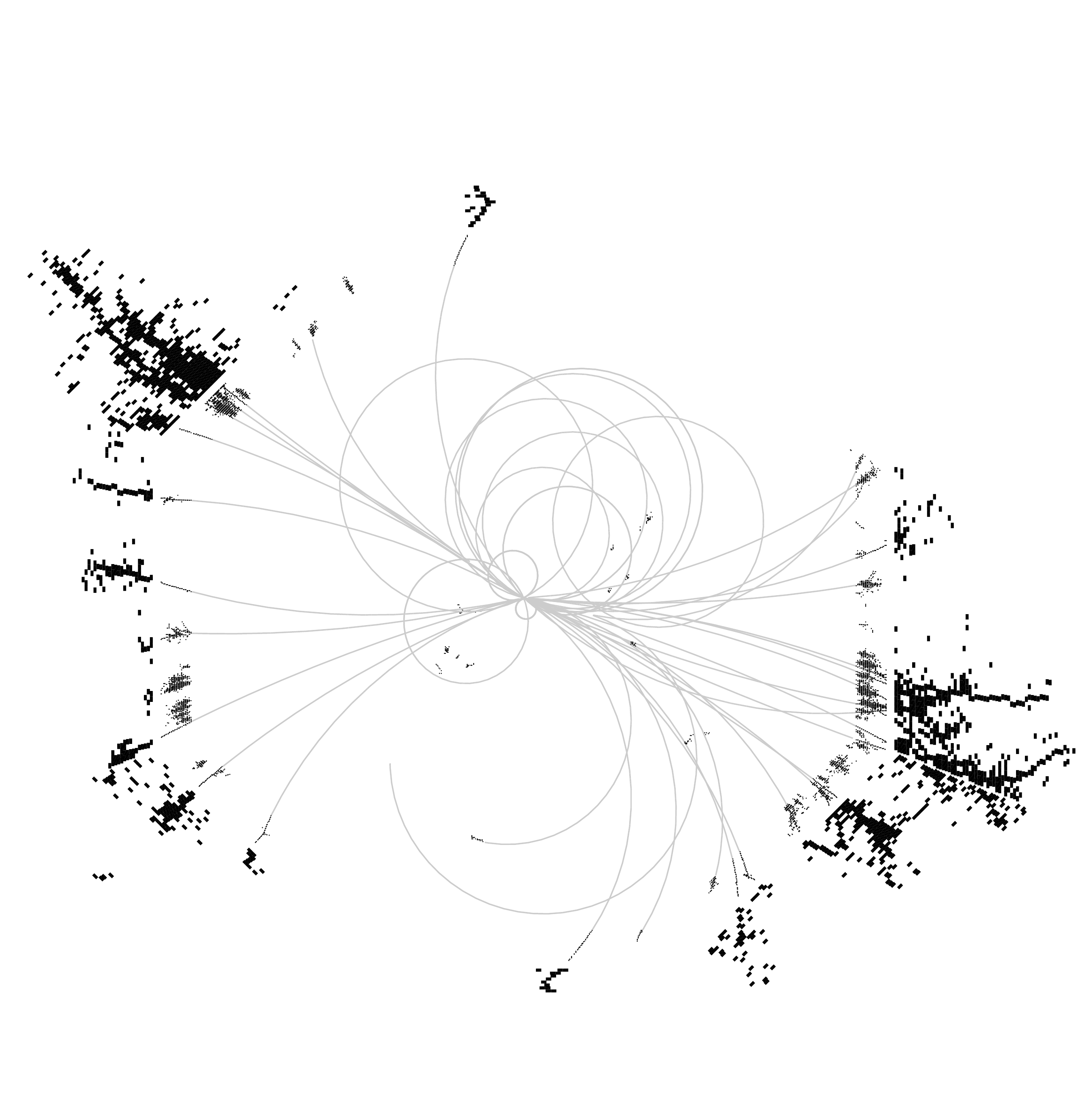}
     \label{fig::ExampleProblemLC}}\\
     \subfloat[][]{\includegraphics[width=0.45\textwidth]{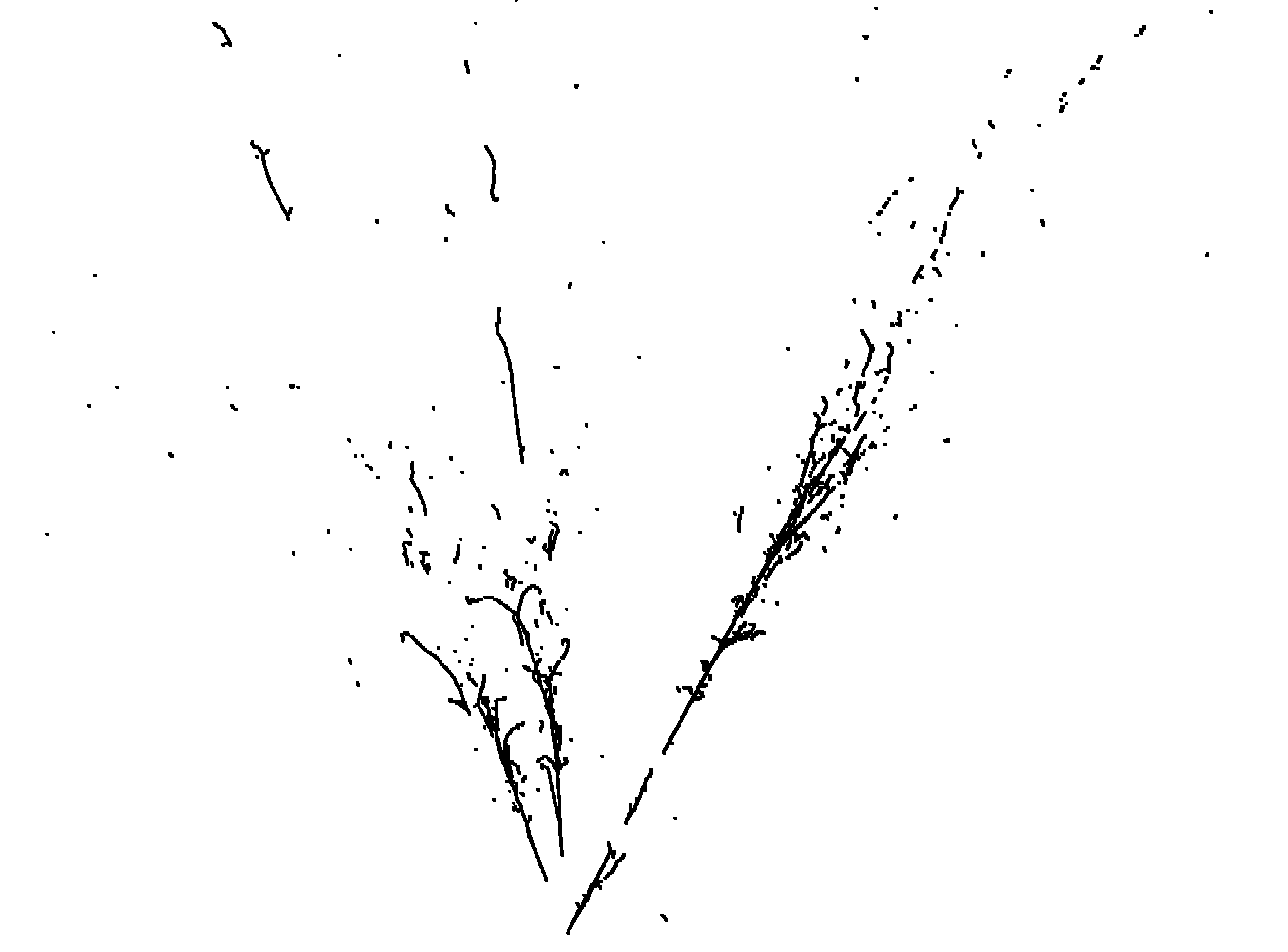}
     \label{fig::ExampleProblemLAr}}
     \caption{\label{fig::ExampleProblems}Typical pattern recognition problems in HEP. (a) Simulated detector response to a Higgsstrahlung event at CLIC. (b) Simulated detector
     response to a charged current $\nu_{e}$ interaction in a LAr TPC.}
  \end{center}
\end{figure}

Figure \ref{fig::ExampleProblems} illustrates two typical pattern recognition problems in HEP. Figure \ref{fig::ExampleProblemLC} shows the simulated detector response to the
production and hadronic decay of Higgs and Z bosons following high energy $e^{+}e^{-}$ collisions at the Compact Linear Collider (CLIC). In order to extract measurements of the Higgs 
boson properties, such as its coupling strengths, it is vital to reconstruct and classify the individual particles in large samples of events. Figure \ref{fig::ExampleProblemLAr}
shows the simulated response of a Liquid Argon Time Projection Chamber (LAr TPC) to a charged current electron neutrino interaction. In order to understand neutrino mixing and
CP-violation in the neutrino sector, it is crucial to identify and characterise each particle in this challenging topology.

The idea underpinning the Pandora SDK is that the interfaces for pattern recognition problems are well defined, as are the operations that must be performed by pattern recognition
algorithms. Whoever poses the pattern recognition problem must specify the building-blocks, or space-points, that define the problem. They must also be able to extract the output
structures, such as clusters, that represent the solution. The algorithms that address the problem must be able to build clusters of space-points and should be able to manipulate
clusters by splitting them up or merging them together. What differs between pattern recognition problems is the precise logic controlling the algorithm operations.

%%%%%%%%%%%%%%%%%%%%%%%%%%%%%%%%%%%%%%%%%

\section{Historical Context}
\label{sec::History}

The Pandora project began in 2007 to provide the first particle flow calorimetry implementation for the proposed International Linear Collider (ILC). A particle flow algorithm was
developed, exploiting the fine granularity detectors in order to reconstruct the paths of individual visible particles. Successful identification of the trajectories allows particle
four-momenta to be extracted from the subdetector system in which they are best-measured, delivering unprecedented jet energy resolution. The Pandora algorithm was used to perform
the first systematic study of the potential of this approach to calorimetry at a high energy lepton collider\,\cite{bib::PandoraNIMA}.

The original Pandora algorithm demonstrated sophisticated pattern recognition ideas, but, in software-engineering terms, was only a proof-of-principle implementation. It was decided
to develop a fully-featured software framework for pattern recognition algorithms and to reimplement the ILC particle flow approach in this new framework. This significant 
software-engineering project took place in 2009-2010 and resulted in the first versions of the Pandora SDK and Pandora Linear Collider content library.

New algorithms were subsequently added to extend the pattern recognition functionality to higher energies, such as those relevant to the multi-TeV lepton collider, CLIC. Pandora was
then used to provide the event reconstruction for the physics analyses described in the ILC Technical Design Report\,\cite{bib::ILC_TDR_2,bib::ILC_TDR_4} and the CLIC Conceptual Design
Report\,\cite{bib::CLIC_CDR_2,bib::CLIC_CDR_3}. The performance of particle flow calorimetry at CLIC was characterised in detail\,\cite{bib::CLICPandoraNIMA}.

The Pandora SDK was designed to be applicable to multiple pattern recognition problems. Most recently, in 2013-2015, a new library of Pandora algorithms was developed to address the
problem of particle reconstruction in the challenging event topologies seen in LAr TPCs\,\cite{bib::LBNEScienceDoc}. This problem is very different to that originally tackled for the ILC, but 
the functionality required from the pattern recognition software framework remains exactly the same.

%%%%%%%%%%%%%%%%%%%%%%%%%%%%%%%%%%%%%%%%%

\section{Overview of the Pandora SDK}
\label{sec::Overview}

The Pandora SDK aims to provide a robust, reliable and easy-to-use environment for developing and running pattern recognition algorithms. Its Application Programming Interfaces
(APIs) are designed to create an environment in which:

\begin{itemize}
    \item It is easy for users to provide the building-blocks defining a pattern recognition problem.
    \item The logic required to solve pattern recognition problems is cleanly implemented in algorithms.
    \item All operations to access or modify building-blocks, or to create new structures, are requested by algorithms and performed by the Pandora framework.
\end{itemize}

This design strategy is well-suited to an approach using large numbers of decoupled algorithms, each of which carefully address specific event topologies, typically controlling
the merging or splitting of clusters.

The Pandora SDK consists of a dependency-free C++ library and carefully-designed APIs. It provides a comprehensive Event Data Model (EDM) for managing pattern recognition problems.
Instances of objects in the EDM are owned by Pandora Manager classes. The instances are stored in named lists and the managers are able to create new objects, delete objects,
create and save new lists and move objects between lists. They provide a complete set of low-level operations that allow high-level operations requested by pattern recognition
algorithms to be satisfied.

\begin{figure}[!h]
  \begin{center}
     \includegraphics[width=0.45\textwidth]{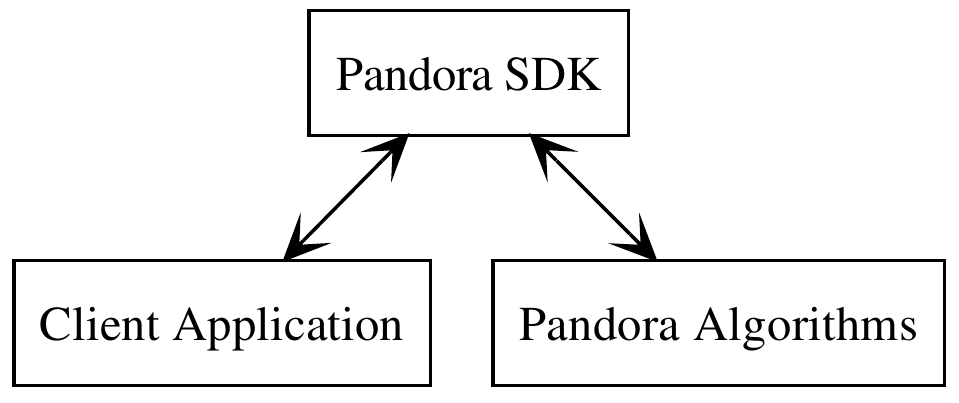}
     \caption{\label{fig::PandoraSetup}The software setup for addressing pattern recognition problems using the Pandora SDK.}
  \end{center}
\end{figure}

To use the Pandora SDK, a user must create a Pandora client application. This provides the input building-blocks to describe the pattern recognition problem and receives the
final output. The pattern-recognition logic is implemented by Pandora algorithms, which ask the Pandora SDK to provide services in order to create new objects or make any
changes to existing instances. Sophisticated visualisation and tree-writing monitoring functionality is available for use by algorithms. Figure \ref{fig::PandoraSetup} illustrates
the typical setup for addressing pattern recognition problems with the Pandora SDK. With this setup in mind, this document will describe the key aspects of the Pandora SDK in detail.

%%%%%%%%%%%%%%%%%%%%%%%%%%%%%%%%%%%%%%%%%

\section{Pandora Event Data Model}
\label{sec::EDM}

The Pandora EDM provides a mechanism for managing data describing pattern recognition problems and their possible solutions. It consists of a set of
classes representing the input building-blocks for a problem and the structures that can be created using these building-blocks. A successful EDM provides a well-defined
development environment for pattern recognition algorithms. It also allows for independence of the algorithms, which can only communicate via the EDM. Algorithms are then
successfully encapsulated and can be developed and maintained independently. An algorithm can be implemented to merge together clusters in close proximity, for instance,
without needing to know anything concerning the construction of the clusters.

The Pandora EDM aims to be self-describing, which is to say that each object provides all the information required to allow investigation and processing by pattern recognition
algorithms. This enables Pandora to be a reusable software solution, completely isolating the pattern recognition algorithms from the details of the software framework and I/O
mechanism used to create or read the input building-blocks.

The building-blocks for pattern recognition in the Pandora SDK are as described below. These are Pandora ``Input Objects" and are typically all created by the
Pandora client application before the pattern recognition algorithms are called (see Section \ref{sec::ClientApp}). These objects are completely defined when they are created
and their properties cannot be changed by the algorithms. The objects are instead used to build new constructs, termed ``Algorithm Objects". The Pandora SDK monitors the usage
of all the Input Objects to ensure that no double-counting can occur, with no Input Object being used to create multiple Algorithm Objects.

\begin{itemize}
   \item \textbf{CaloHit} The primary building-block for pattern recognition problems, a CaloHit defines a position and extent in space and time, together with an associated
   intensity or energy measurement. Whilst CaloHits can represent points in free space, they can also provide information regarding their location in a particle detector.
   This includes details of the subdetector system in which energy was deposited and information about the calorimeter readout and geometry. The CaloHits hold estimators for the
   electromagnetic energy or hadronic energy associated with the space-point. It is for algorithms to select the appropriate energy estimator.
   \item \textbf{Track} Continuous trajectories of well-defined space-points are represented by Track objects. These are helix parameterisations of the space-points, providing
   details of particle positions and momenta (Track States) along the trajectory. These objects were originally designed to represent Tracks reconstructed in
   fine granularity, low material-budget tracking systems in particle detectors. As such, key information provided by Pandora Track objects include impact parameter details and
   a projected Track State at the surface of the detector calorimeters. Tracks can have parent-daughter and sibling relationships in order to fully-describe particle interactions
   and decays that can occur within a tracking detector.
   \item \textbf{MCParticle} Primarily for development purposes, MCParticles can be provided for access by the pattern recognition algorithms. These provide full
   details of the true pattern recognition solution for simulated events. MCParticle instances can have parent-daughter links and can fully describe particle decay cascades in
   simulated interactions. The MCParticles can store details of their association (in terms of e.g. true energy deposited) with each CaloHit and Track. Using MCParticles, it
   is possible for algorithms to cheat some, or all, aspects of the pattern recognition, allowing a wealth of development and debugging functionality.
\end{itemize}

The Pandora Algorithm Objects represent the higher-level structures created in order to solve pattern recognition problems. The Pandora SDK carefully manages the allocation
and manipulation of these objects and all non-const operations can only be requested by algorithms via the Pandora APIs. The Pandora SDK is then able to perform the
memory-management for the objects.

\begin{itemize}
   \item \textbf{Cluster} The main working-horse for pattern recognition algorithms, a Cluster is a collection of CaloHits. It also provides derived information describing
   the combined properties of the CaloHit collection, such as energy estimators and the results of linear fits to the CaloHit spatial positions. The most typical tasks for
   Pandora algorithms will be to create new Clusters from lists of input CaloHits or to read lists of input Clusters and selectively split or merge some Clusters.
   \item \textbf{Vertex} The identification and classification of a specific point in space, vertices are typically used to flag positions of particle creation or decay.
   \item \textbf{ParticleFlowObject} A container of Clusters, Tracks and Vertices, together with metadata describing the particle type and four-momentum. The Particle Flow
   Object (PFO) is the ultimate output of the pattern recognition, grouping the input objects into structures that completely define the solution. PFOs can have parent-daughter
   links in order to describe particle decay hierarchies.
\end{itemize}

Instantiation of objects in the Pandora EDM follows a design pattern that provides a clean and simple interface. Object creation is typically requested by a client application
(Input Objects) or an algorithm (Algorithm Objects). The calling function must create a local instance of a Pandora parameters object. There is one parameters class for each
type of Pandora object, each having public member variables to which values must be assigned. For instance, a PandoraApi::CaloHit::Parameters instance will have public member
variables such as the CaloHit position (three-vector) and dimensions. The calling function must assign to each of the member variables, then call the Create API.
Failure to assign to any member variable in the parameters instance will lead to failure of the object creation. Successful assignment will allow the Pandora SDK to create
the required object instance. The newly created instance is owned and managed by the Pandora SDK, but can be accessed and manipulated by algorithms, as described in
Section \ref{sec::Managers}.

It is possible, for advanced users of the Pandora SDK, to add additional functionality to the base objects in the Pandora EDM. The user can inherit from the Pandora base class
for the relevant object. The user may also want to add additional content to the relevant parameters class and can also define a new class inheriting from the base parameters.
The key step is for the user to provide an object instantiation factory, inheriting from the ObjectFactory template base class, to perform the actual object creation. The
Pandora SDK will continue to work with pointers to the base class for these derived objects, but algorithms can successfully access additional functionality via use of a dynamic
cast.

The Pandora object creation mechanics are templated to allow for simple addition of new objects to the EDM. Such an addition would require only the definition of a new object
type, plus the associated parameters class.

%%%%%%%%%%%%%%%%%%%%%%%%%%%%%%%%%%%%%%%%%

\section{Pandora Client Application}
\label{sec::ClientApp}

\begin{figure}[!h]
  \begin{center}
     \includegraphics[width=0.29\textwidth]{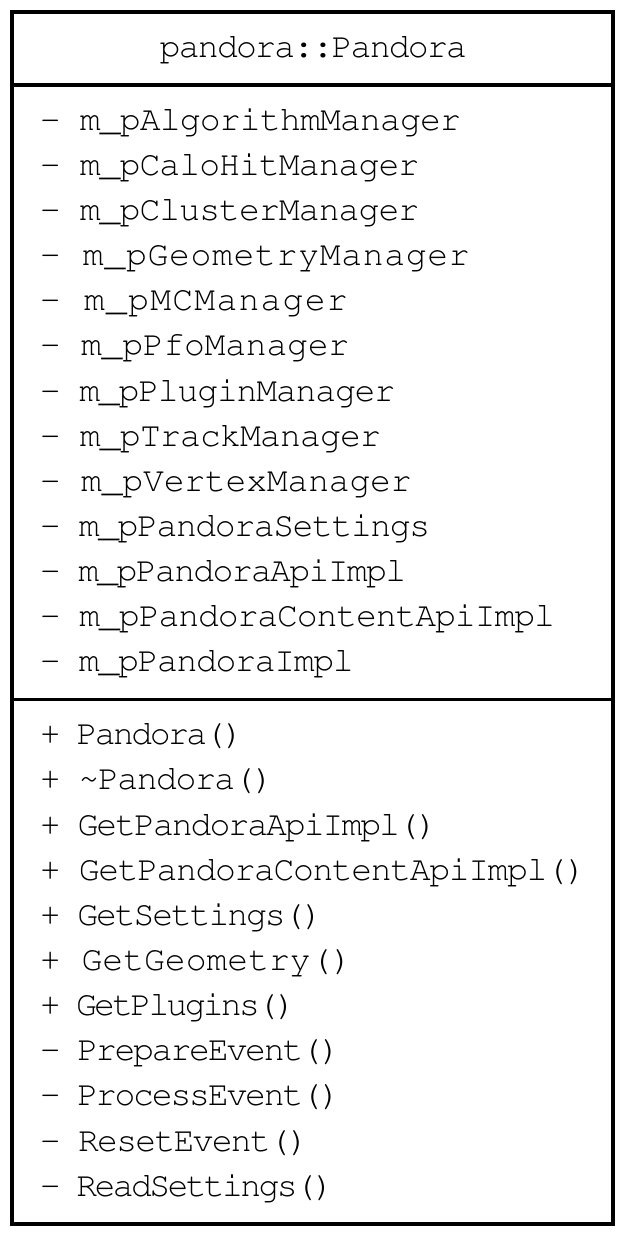}
     \caption{\label{fig::PandoraInstance}Class diagram describing the structure of the Pandora class. The member variables consist of the addresses of Pandora manager instances,
     API implementation instances and a settings instance. The member functions provide a number of high-level services and are typically accessed via the Pandora APIs.}
  \end{center}
\end{figure}

The Pandora client application is ultimately responsible for controlling pattern recognition reconstruction using the Pandora SDK. The client application creates the
Pandora instances and then sends requests to these instances. During the initialisation step, it must use the Pandora APIs to perform the following operations:

\begin{enumerate}
    \item Create the required Pandora instances. Typically only a single instance is required, although advanced use-cases addressing problems where particles are split
    between multiple detectors may require more. A Pandora instance, as shown in Figure \ref{fig::PandoraInstance}, contains instantiations of the Pandora manager classes
    and API implementations.
    \item Register the required Algorithm and Algorithm Tool Factories with the Pandora instance(s). These factories give a Pandora instance the ability to create
    instances of the algorithms and algorithm tools (see Section \ref{sec::AlgorithmTools}), if they are requested via the PandoraSettings XML configuration file. 
    \item Ask the Pandora instance(s) to parse the provided PandoraSettings XML files, which describe the chain of pattern recognition algorithms to be used
    to process each event. The Pandora instance will create and manage the algorithm instances as required and will configure them as specified.
\end{enumerate}

On a per-event basis, the client application must perform the following operations:

\begin{enumerate}
    \item Ask the Pandora instance(s) to create the building-blocks for the pattern recognition problem. As described in Section \ref{sec::EDM}, this involves assigning
    values to each of the fields required for the relevant object type. The Pandora algorithms access the information stored in these building-blocks, but, crucially,
    do not need to know how the information has been obtained. The client application thus isolates algorithms from the user's software framework.
    \item Ask each Pandora instance to process the event. The thread will be passed to the Pandora instance, which will then run the specified algorithms in the specified
    order, processing the requested building-blocks in order to produce output Clusters, Vertices and PFOs.
    \item After algorithm processing, ask each Pandora instance to provide its solution to the pattern recognition problem. This will typically be in the form of PFOs. The
    client application can read the list of reconstructed PFOs and access the building-blocks that form each PFO. By using the parent addresses/identifiers specified
    for each building-block, the user can link the PFO constituents back to objects in the input software framework. The user will typically want to persist the output.
    \item Ask to reset each Pandora instance, in preparation for the next event. This will reset all of the Pandora manager classes so that all Input Objects and Algorithm 
    Objects are removed and all saved object lists are removed/reset.
\end{enumerate}

The client application is thus responsible for defining the pattern-recognition problem and persisting the solution. It is also responsible for bringing together algorithm
implementations and for configuring the Pandora instances. Algorithms can, for convenience, be bundled together into Pandora Content Libraries. The client application can then
simply ask to register factories for all the available algorithms in a given content library. The algorithms will depend on the Pandora SDK, but can also have as many external
dependencies as required by their implementation. The client application will depend on the Pandora SDK and on the content libraries. The actual algorithm instances used in the
reconstruction are not created until the Pandora instances parse the PandoraSettings XML file.

%%%%%%%%%%%%%%%%%%%%%%%%%%%%%%%%%%%%%%%%%

\section{Pandora Managers}
\label{sec::Managers}

The aim of the Pandora SDK is to provide key services for pattern recognition algorithms, so that the algorithms can remain simple and focused on pattern recognition logic.
At the heart of this design are the Pandora manager classes, which own all instances of objects in the Pandora EDM. The managers aim to provide a complete set of low-level
object manipulation functions. Algorithms request high-level services, which are then satisfied when the API implementations, or the managers themselves, call the correct
low-level manager functions in the correct order. This approach helps to ensure that the implementation is extensible, easy to maintain and rather human-readable. A key part
of the design is that algorithms can \textit{only} access managed objects via the Pandora APIs, so the managers are able to perform memory-management and book-keeping.

\begin{figure}[!h]
  \begin{center}
     \hspace*{-0.95 cm}
     \includegraphics[width=0.55\textwidth]{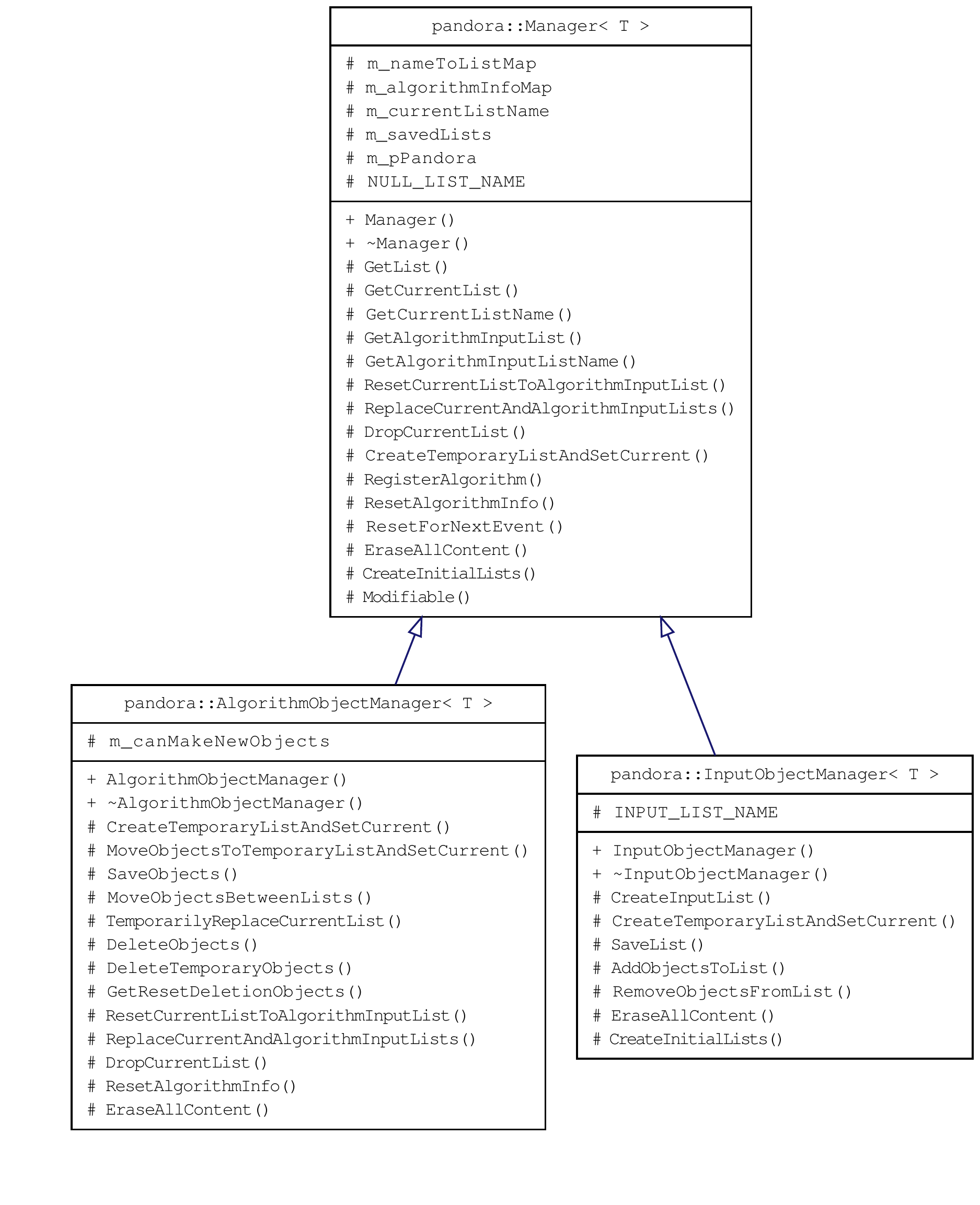}
     \caption{\label{fig::Managers}Class diagram describing the Manager template base classes. The Manager class provides functionality for supervising named lists of
     objects. The derived classes provide functionality that reflects the different rules governing creation and usage of Algorithm Objects and Input Objects.
     Concrete object manager classes derive from the appropriate base classes.}
  \end{center}
\end{figure}

The Pandora manager classes are templated on the managed-object type, allowing for easy addition of new types to the Pandora EDM. There is a Manager template base class and
separate derived template classes for Input Object and Algorithm Object Managers. There are also manager classes for each of the object types in the Pandora EDM, which derive from 
the appropriate base classes. These address small details specific to each object type. For instance, the Track Manager handles track parent-daughter and sibling 
relationships. A Pandora instance owns manager instances for each of the object types in the EDM. The structure of the manager classes is illustrated in Figure \ref{fig::Managers}.

Pandora objects are heap-allocated and their addresses are stored in named object lists, owned by the object managers. The object lists are unordered sets, keyed on pointers
to the objects. This storage strategy ensures efficient retrieval of specific object instances, although care is required if ever the objects must be sorted in a defined manner.
Each manager holds a map from the list name (a string) to the address of the object list. It also stores the set of saved list names, plus the name of the \textit{current} list.

Algorithms can use the Pandora APIs in order to receive const references to the object lists from the managers. Algorithms can access lists by name, or can simply ask
for the current list. Algorithms can also choose to replace the current list name with the name of another saved object list. This aids the development of reusable algorithms
where, for instance, parent algorithms can control the current lists before running instances of daughter algorithms to process the contents of the current lists.

The manager classes all hold the address of the Pandora instance with which they are associated. They also record details for all the algorithms that are currently running, such
as the current list name when the algorithm was first called and details about any temporary lists the algorithm has created. Note that the algorithm stack may have more than
one entry, since parent algorithms can run daughter algorithms. The details of accessing object lists are common to all managers, but details concerning the creation of 
new object lists and saving lists are different for Input Objects and Algorithm Objects.

\subsection{Input Object Managers}
Input Objects can be created, via the Pandora APIs, by any function with access to the relevant Pandora instance. The most common point of creation is, however, a Pandora client
application. Newly-requested objects are created on the heap, via the relevant manager, and their address is always stored in a specific named list: the ``Input" list.

The design idea is that Input Objects cannot be modified or deleted by the pattern recognition algorithms, although new refined or reimagined objects could be created by
algorithms. The Input list, for a given manager, keeps a full record of all the objects created. Algorithms can choose to work with this list, or, more typically, save new
lists (under new names) containing only a subset of the Input lists. Copies of the pointers to Input Objects can appear in multiple saved lists.

The memory-management is rather simple, as all Input Objects are deleted only when the client application asks to reset the managers at the end of an event. All objects in the
Input list can be deleted and all other lists simply cleared and deleted.

\subsection{Algorithm Object Managers}
Memory-management for Algorithm Objects is considerably more complex than for Input Objects. Algorithm Objects will typically be created, modified and deleted as the pattern
recognition reconstruction progresses. The Pandora SDK enforces a specific approach to working with Algorithm Objects, which maintains flexibility whilst also enabling use of
the powerful reclustering functionality described in \mbox{Section \ref{sec::Reclustering}}.

In order to create a new Algorithm Object, the relevant manager must have a new, temporary object list as the current list, waiting to receive newly created instances. The
creation of such a temporary list must be requested by an algorithm. If the temporary object list is not in-place, the object creation will be rejected. Similarly, object
creation is not possible once the algorithm finishes or chooses to run a daughter algorithm. If the temporary list is present, it will receive the address of any newly-created
objects. The temporary list is associated with the algorithm that requested its creation. When this algorithm ceases running, its associated temporary lists will be removed and
all Algorithm Objects contained will be deleted. In order to persist the Algorithm Objects, the algorithm must ask to save the objects in a (new or existing) named object list.

Once saved in a named list, additional APIs are available to move (subsets of) objects between existing lists or to new lists. Unlike Input Objects, it is enforced that the address
of each Algorithm Object can exist in only one list. At the end of an event, Algorithm Objects are tidied-up by deleting all instances in all saved lists and then deleting all the
lists. Any Algorithm Objects that were created, but not explicitly saved, are deleted automatically upon algorithm completion.

The mechanics for Algorithm Object management were designed with Pandora reclustering (Section \ref{sec::Reclustering}) firmly in mind. These mechanics keep the
Algorithm Objects under extremely tight control. It could be argued that all objects in the Pandora EDM should behave as Algorithm Objects, and the implementation to achieve
this would be trivial as all managers and APIs are templated. The existing design decision, however, is simply to allow more flexibility for the Input Objects, where
memory-management can be achieved rather simply.

\subsection{Monitoring of Object Usage}
The Algorithm Objects are typically containers of other objects. Clusters, for instance, are containers of CaloHits, whilst PFOs are containers of Clusters, Tracks and Vertices.
An important role played by the manager classes is the monitoring of object usage, ensuring that no double-counting can occur. When an algorithm asks to create an Algorithm Object,
or add to an Algorithm Object, the availability of the daughter objects is checked. If any of the daughter objects is flagged as unavailable, the Algorithm Object creation or 
modification will not be allowed. If the daughter objects are available, the operation can proceed and the daughter objects will be flagged as unavailable. Removal of the daughter
objects will reset their availability flags. Algorithms can use the Pandora APIs to ask whether objects are available or have already been used, functionality which proves
extremely useful.

The availability monitoring is largely straightforward, with each instance of a daughter object having a boolean member variable that can only be modified by the relevant manager
class. For CaloHits the operation is, however, complicated significantly by the Pandora reclustering mechanism. This mechanism is described in detail in Section \ref{sec::Reclustering},
and allows multiple Cluster configurations to be explored simultaneously. The CaloHit manager performs CaloHit usage monitoring separately for each set of Cluster configurations.
It ensures that a CaloHit can appear simultaneously in multiple Cluster candidates during reclustering, but can only feature once in the final selected Clusters.

%%%%%%%%%%%%%%%%%%%%%%%%%%%%%%%%%%%%%%%%%

\section{Pandora Algorithms}
\label{sec::Algorithms}

Pandora algorithms contain the step-by-step instructions for finding patterns in the provided data. The algorithms use the Pandora APIs to access the Pandora objects and
to request the Pandora managers to make new objects or modify existing objects. Algorithms inherit from the Pandora Process purely abstract base class. The inherited functionality
controls the handshaking procedure between the Pandora instance and the algorithm instance, establishing a communication channel between the two entities. The Process base class
also provides the ability to receive a ReadSettings callback with a provided XML handle from which any configurable parameters can be extracted. Finally, Process instances can also
receive an Initialise callback, providing an opportunity to finalise any constructs required before event processing can begin. The Algorithm purely abstract base class then provides
the interface for the crucial Run callback, which is called each event and is the entry-point for all event processing.

\subsection{Algorithm Creation and Configuration}
As discussed in Section \ref{sec::ClientApp}, the Pandora client application is responsible for registering Algorithm Factories with a Pandora instance. Algorithm factories are
extremely simple, each implementing a CreateAlgorithm function that allocates an instance of the relevant derived algorithm class and returns a pointer to the Algorithm base class.
Algorithm factories are registered with the Pandora Algorithm Manager under specific names. Pandora is configured by an XML file, the path to which is provided by the client
application. When Pandora parses the PandoraSettings XML file, it will look for algorithm XML tags within the top-level Pandora XML tags. For each algorithm XML tag found, Pandora
will extract the algorithm type, which must match the name of a registered algorithm factory. If there is a match, the Algorithm Manager will ask the factory to create a new instance
of the desired algorithm type and will store the pointer to the Algorithm base class.

After algorithm creation, the Algorithm Manager will call the ReadSettings member function of the new algorithm, providing a handle to the XML element describing the algorithm.
This provides the mechanism by which Pandora algorithms can be configured at run-time. The ReadSettings function can look for specific daughter XML tags and assign values to its
member variables as required. The algorithm can demand that some XML tags be present, returning a context-specific error to halt execution if configuration details are missing.
Alternatively, each algorithm can assign default values to each of its configurable member variables, then provide the ability to override the default values if specific tags are
found in the XML file. This allows an implementation in which algorithms need have no hard-coded parameters, but the XML configuration can remain clean and simple for most use-cases.

The Algorithm Manager will create and configure instances of all the algorithms specified in the top-level of the PandoraSettings file. It will store the
pointers to the Algorithm base classes in an ordered container. It will call the Initialise method for each algorithm, then is ready for event processing. When the client
application asks the Pandora instance to process an event, the Algorithm Manager will call the Run method for each of the algorithms in turn. Upon completion of the final algorithm
in the ordered list, the thread will be returned to the client application.

\subsection{Nested Algorithms}
The Pandora Algorithm Manager will only search for algorithm XML tags within the top-level Pandora XML tags. These are the algorithms that it will create, configure, initialise
and then run, in order, each event. In the configuration step, however, when a new algorithm receives a callback to its ReadSettings method, the algorithm itself is given control
of parsing any details contained within its XML tag. The algorithm can then search for daughter algorithm XML tags. Daughter algorithms could be specified in a list, within a named
XML tag, or may be identified by a XML description attribute. If found, the parent algorithm can use an API to instruct the Algorithm Manager to create a new instance of the
specified algorithm type, configure the new algorithm instance and then return the unique name of the new algorithm instance. During event processing, the parent algorithm can
use an API to ask to run the daughter algorithm instance with the stored unique name. This nesting of algorithms opens-up functionality whereby parent algorithms can manipulate
the current object lists, for instance, then call reusable daughter algorithms to process the objects in the current lists.

\subsection{Algorithm Tools}
\label{sec::AlgorithmTools}
Using nested daughter algorithms to perform operations promotes the creation of small, reusable algorithms containing just the kernel of specific pattern-recognition logic.
The design does mean, however, that the parent and daughter algorithms are completely decoupled and there can be no communication between the two, other than via manipulation
of objects in the EDM or the object lists. Sometimes, the user will instead want to create Algorithm Tools, which can offer direct extensions to specific algorithms.

\begin{figure}[!h]
  \begin{center}
     \includegraphics[width=0.45\textwidth]{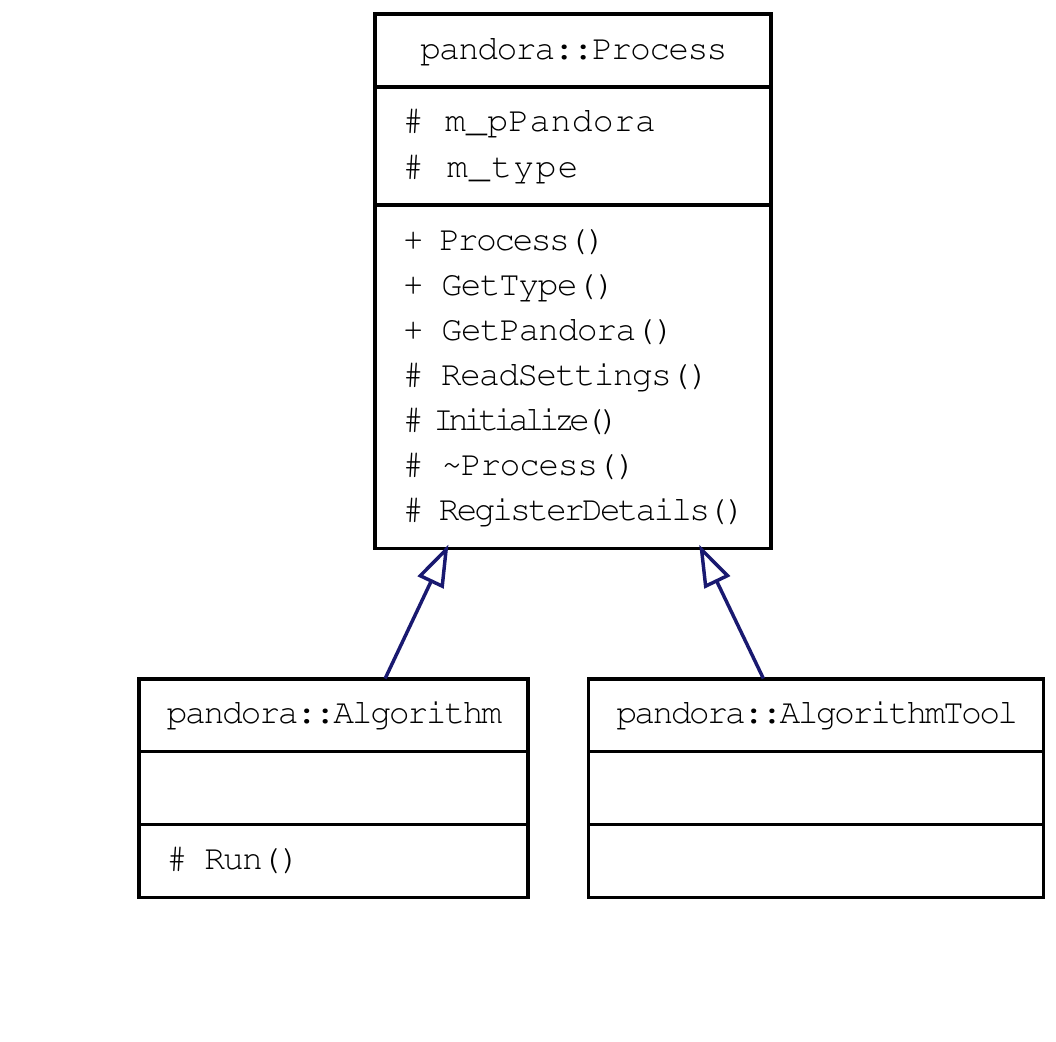}
     \caption{\label{fig::Algorithms}Class diagram describing the structure of the Algorithm and Algorithm Tool classes. The Process base class provides all functionality for
     handshaking with a Pandora instance and for XML-based configuration. Algorithms must provide an implementation of the Run method, which is their entry-point for event
     processing. Algorithm Tools have a user-defined interface and provide custom functionality to parent algorithms.}
  \end{center}
\end{figure}

An algorithm tool inherits from the Pandora Process purely abstract base class, so it has all the handshaking and configuration functionality of an algorithm. It does not, however,
receive a Run callback from the Algorithm Manager. Instead, a parent algorithm defines the interface for its algorithm tools and is given direct access to the pointer to the algorithm
tool instances. The parent algorithm can create some complicated construct (e.g. comparing multiple Clusters), then hand the construct directly to its algorithm tools for
processing. The precise algorithm tools used can be specified via the PandoraSettings XML file, allowing for simple run-time configuration. The structure of the algorithm and
algorithm tool classes is illustrated in Figure \ref{fig::Algorithms}.

%%%%%%%%%%%%%%%%%%%%%%%%%%%%%%%%%%%%%%%%%

\section{Pandora APIs}
The APIs are static functions, which are typically templated to allow for operations on each of the different types in the Pandora EDM. The APIs used by the Pandora algorithms,
for accessing and manipulating Pandora content, take a reference to a Pandora algorithm as their first argument. This allows the static functions to resolve a particular Pandora
instance and access the relevant instance of the API implementation. Careful friending of classes ensures that the API implementation is in the privileged position
of being able to call Pandora manager functionality inaccessible to other classes. The API implementation typically calls manager functions directly, but sometimes provides simple
logic, calling multiple functions in different managers in order to provide the high-level services requested by algorithms. The APIs used by the client application instead take
a reference to a Pandora instance as their first argument, but otherwise work in an identical manner.

Typical uses of the Pandora APIs are demonstrated in Algorithms \ref{alg::ExampleAlg1} and \ref{alg::ExampleAlg2}, which describe, in human-readable form, the API calls that are
required in order to perform key operations. Algorithm \ref{alg::ExampleAlg1} illustrates access to a list of CaloHits, followed by the creation of Clusters, which are saved in a
named list. Algorithm \ref{alg::ExampleAlg2} shows the merging of pairs of Clusters. In each case, the true implementation in a Pandora algorithm will read in exactly the same
manner, with the addition of encapsulated logic and decision-making procedures to determine exactly when to create a Cluster or merge Clusters. A comprehensive set of demonstrations
and unit-tests are included in the Pandora ExampleContent library and test application, which are described in\,\cite{bib::ExampleContent}.

\begin{algorithm}[h]
  \begin{algorithmic}[1]
    \Procedure{Cluster Creation}{}
      \State Create temporary Cluster list 
      \State Get current CaloHit list
      \ForAll{CaloHits}
	\If{CaloHit available}
          \ForAll{newly-created Clusters}	
            \State Find best host Cluster
          \EndFor
	  \If{Suitable host Cluster found}
	    \State Add CaloHit to host Cluster
	  \Else
	    \State Add CaloHit to a new Cluster
	  \EndIf
	\EndIf
      \EndFor
      \State Save new Clusters in a named list
    \EndProcedure
  \end{algorithmic}
  \caption{Cluster creation pseudocode. The logic determining when to create new Clusters and when to extend existing Clusters will vary between algorithms.}\label{alg::ExampleAlg1}
\end{algorithm}

\begin{algorithm}[h]
  \begin{algorithmic}[1]
    \Procedure{Cluster Merging}{}
      \State Get current Cluster list
      \ForAll{Clusters}
        \If{Cluster is suitable parent}
	  \ForAll{Clusters}
	    \State Find best daughter Cluster
	  \EndFor   
	  \If{Suitable daughter Cluster found}
	    \State Merge daughter Cluster into Parent
	  \EndIf
	\EndIf
      \EndFor      
    \EndProcedure
  \end{algorithmic}
  \caption{Cluster merging pseudocode. The logic governing the identification of suitable parent Clusters and daughter Clusters will vary between algorithms.}\label{alg::ExampleAlg2}
\end{algorithm}

%%%%%%%%%%%%%%%%%%%%%%%%%%%%%%%%%%%%%%%%%

\section{Pandora Reclustering}
\label{sec::Reclustering}

The Pandora reclustering mechanism exploits the functionality of the Pandora SDK in order to reinvent the traditional pattern recognition approach. Using reclustering allows
algorithms to simultaneously explore multiple different Cluster configurations. The resulting Clusters can be compared and a decision made as to which configuration is best.
The Pandora SDK will then automatically tidy-up any discarded Clusters. This approach means that, instead of selecting the best algorithmic approach to solve a problem, the
user is able to control a process whereby the approach that best \textit{solved} the problem is identified. This allows use of the Pandora XML configuration to drop-in a
(potentially large) number of independent approaches to solving a problem. The overall algorithm steering will then select the approach that is deemed most successful.

\subsection{Standard Reclustering}
The standard reclustering use-case is to examine the Clusters produced by a clustering algorithm and identify problems or deficiencies with a subset of these Clusters.
For example, in the Pandora Linear Collider event reconstruction, a significant discrepancy between measured inner detector Track momentum and associated calorimeter Cluster
energy indicates a pattern recognition failure. A Pandora algorithm identifying such a failure can use an API to ask to Recluster any Clusters (and Tracks) in the current object
lists. This request moves the relevant Clusters from the current list into a new temporary Cluster list. Furthermore, CaloHits in the relevant Clusters are copied into a new
CaloHit list, which is set as the current CaloHit list. The Tracks are copied into a new Track list, set as the current Track list.

The steering algorithm can then ask to run a daughter clustering algorithm. A new temporary Cluster list is created to hold any newly-created Cluster candidates. The temporary
list is associated with the steering algorithm, which receives the temporary list name. The daughter clustering algorithm can access the current CaloHit and Track lists and create
an all-new Cluster configuration, stored in the temporary list. The steering algorithm can repeatedly call daughter clustering algorithms, using differently-configured instances
of the same algorithm or radically different clustering approaches. The result is a set of named temporary Cluster lists, each holding Cluster candidates. One named temporary
list holds the original, input Clusters.

By examining the different temporary Cluster lists, the steering algorithm makes a decision as to which Cluster configuration is the best. For the Linear Collider reconstruction
discussed above, this would be the configuration in which Track momentum and associated Cluster energy match most closely. The steering algorithm then only needs to call the
EndReclustering API, specifying the name of the chosen temporary Cluster list. The desired Clusters will be injected into the original current list, replacing those originally present.
All other temporary lists will be cleaned by the Cluster manager, deleting all the rejected Cluster candidates and all the temporary lists. The end result is a seamless replacement
of the original, deficient Clusters using the best result from a large number of alternative (black-box) clustering algorithms.

\subsection{Local Reclustering}
In some use-cases, a steering algorithm may not wish to use daughter clustering algorithms in order to reconfigure existing Clusters. The algorithm may instead want to simply
examine one alternative configuration and to directly compare the original and new Cluster candidates side-by-side. This functionality is provided by a local reclustering
mechanism offered by the Pandora SDK. The steering algorithm must ask to Fragment the original Cluster candidates, which will move the relevant Clusters into a new temporary
Cluster list, plus create a further temporary Cluster list to receive any newly-created Clusters. The same algorithm can now build a second set of Cluster candidates, from
the same constituent CaloHits, without violating the CaloHit usage monitoring requirements. The two sets of Cluster candidates can be compared, before a decision must be made
as to which of the configurations should be selected. The steering algorithm must ask to end the fragmentation process, specifying which of the two temporary Cluster lists to
keep and which to delete. The chosen Clusters will seamlessly replace those in the input current list.

%%%%%%%%%%%%%%%%%%%%%%%%%%%%%%%%%%%%%%%%%

\section{Additional Functionality}
\label{sec::Additional}

In addition to the Pandora EDM and the functionality provided by the Pandora APIs, the Pandora SDK offers a number of other key features to aid pattern recognition algorithm
implementation, development and debugging. These include classes to complement the EDM, providing three-vector algebra or performing (sliding) linear fits to Clusters. The
most important additional features are described in this Section.

\subsection{Plugins}
\label{subsec::Plugins}
The Pandora SDK can be easily extended via the addition of Plugins. These inherit from the Pandora Process purely abstract base class and have interfaces that define their
specific usage. Plugins are currently available to provide a number of services specific to pattern recognition in HEP. These include particle identification, energy correction
functionality and electromagnetic shower profile characterisation. Plugins are also available to provide access to magnetic field maps and to divide regions of a particle
detector into layers. The plugins are owned by a Plugin Manager and can be accessed by algorithms via a GetPlugins API. The plugins inherit from the Process base class
all functionality required for handshaking with the relevant Pandora instance and for configuration via the PandoraSettings XML file.

The particle identification plugins receive a Pandora Cluster and return a boolean result recording whether the Cluster matches the topology expected for a particular particle
type. The energy correction plugins, meanwhile, receive a Pandora Cluster and implement a custom scheme for estimating the energy of the Cluster, returning the energy estimate
via a reference to a float. This plugin functionality allows for improvements whenever it is possible to obtain a better energy estimator using means other than simply
summing the constituent CaloHit energies. This covers digital calorimeters, for instance, which record only whether a particle was detected in a particular readout cell, not an
analogue energy measurement.

\subsection{Detector Geometry}
Specific to HEP is the ability for the client application to provide detector geometry information, which can then be requested by Pandora algorithms. The client application
can specify details of subdetector systems such as the type (e.g. electromagnetic calorimeter), name and spatial extent, where the assumption is that the subdetector will
have a polygonal structure. Details for each of the layers of active material can also be provided, as can information about gap regions in the detector active material.
Algorithms can choose to use the detector properties to help define their pattern recognition logic. In general, however, Pandora algorithms try to avoid use of detector
information and work with the self-describing CaloHits and Tracks alone. The detector geometry information is then used predominantly for event visualisation purposes.

\subsection{Monitoring and Visualisation}
Algorithms can access a Pandora Monitoring library, which has a dependency on the ROOT data analysis framework\,\cite{bib::ROOT}. ROOT provides the ability to write tree
structures, allowing information constructed within algorithms to be recorded and written to file. The trees can store information from large numbers of events and allow algorithm
logic to be developed and tuned via examination of high-statistics distributions. ROOT also offers a Event Visualisation Environment (EVE), which provides an application framework
for constructing event display programs. The Pandora Monitoring library provides the translation from the Pandora EDM to the ROOT EVE data model.

Algorithms can use the Pandora Monitoring APIs to request visualisation of custom lists of any objects in the Pandora EDM, specifying the visualisation colour-scheme and providing
a name to identify the objects in the GUI (where objects can be queried and toggled on or off). A typical approach is for algorithms to request visualisation of multiple lists of
objects, then ask to view the event. The event can be examined in the GUI and the Pandora algorithm processing will be paused until the user hits return in the relevant terminal.
The event display provides zoom and rotation functionality for examining events in three dimensions. Algorithms can also choose to add reference marker points to highlight
positions of particular interest in an event.

Visualisation can aid algorithm development. It is possible for a developer to visually check the results of a Cluster selection procedure, for instance, by asking for separate
displays of all Clusters passing or failing the selection. When looking to merge Clusters, the event display can show all the possible parent and daughter Cluster combinations,
alongside details of their association properties, which can be printed to the terminal. Reusable event display algorithms, which can be included in the PandoraSettings XML file
in multiple locations, can provide visualisation of the pattern recognition progress at different points in the reconstruction, without the need to recompile any source code. This
visual approach to development and debugging can be a particularly efficient and rewarding way to create pattern recognition algorithms.

\subsection{Persistency}
As described so far in this document, a Pandora reconstruction is always controlled by a client application running in the user's chosen software framework. The client application
asks the Pandora SDK to create self-describing objects to define the pattern recognition problem. Pandora persistency provides a means whereby the self-describing building-blocks
can be written-to or read-from files. This allows a user to run a client application just once, calling the Event Writing algorithm, which uses the Pandora persistency APIs to
write the events to binary PNDR files (small, but platform-specific) or to XML files (large, but platform-independent and easily compressed). Subsequently, the user can run in a
minimal Pandora standalone application, using the Event Reading algorithm to recreate the self-describing objects for processing by the algorithms. This functionality allows for
rapid development outside of complex software frameworks, such as those typically used in HEP.

%%%%%%%%%%%%%%%%%%%%%%%%%%%%%%%%%%%%%%%%%

\section{Pandora Reconstruction Examples}
\label{sec::Reconstruction}
The algorithms to solve specific pattern recognition problems can be distributed in Pandora content libraries, which contain all the required algorithms, algorithm tools and plugins.
Client applications can register all products in a content library with a Pandora instance by calling a single function.

This document now brings together all the information from the preceding Sections in order to describe the Pandora solutions to the pattern recognition
problems in Figure \ref{fig::ExampleProblems}. The reconstructions are provided by separate Pandora content libraries. The pattern recognition problems, and so algorithms, are
rather different, but each uses many decoupled algorithms to gradually build particles, trying to avoid mistakes.

\subsection{Linear Collider Event Reconstruction}
\begin{figure}[!h]
  \begin{center}
     \subfloat[][]{\includegraphics[width=0.45\textwidth]{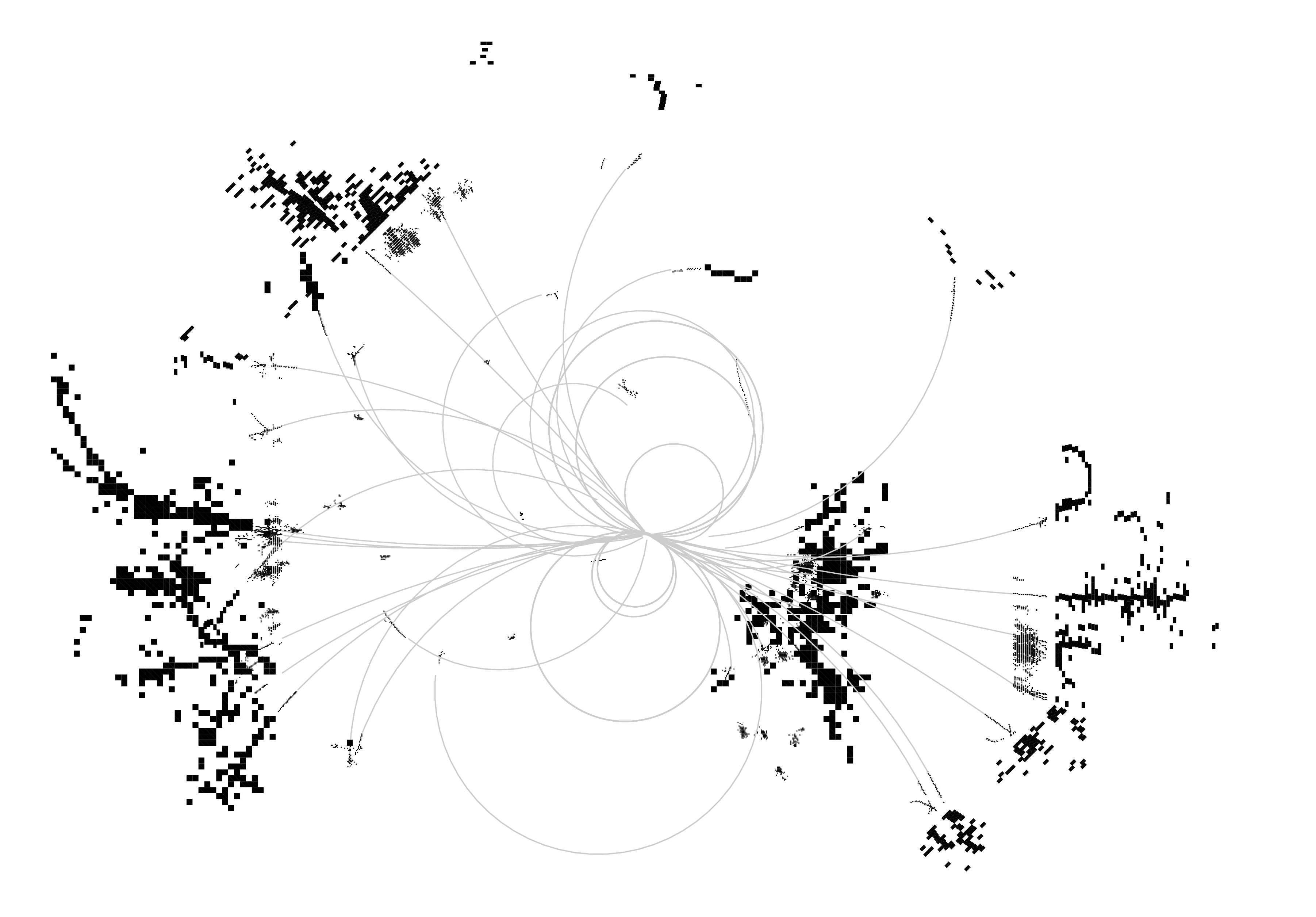}
     \label{fig::LCRecoBefore}}\\
     \subfloat[][]{\includegraphics[width=0.45\textwidth]{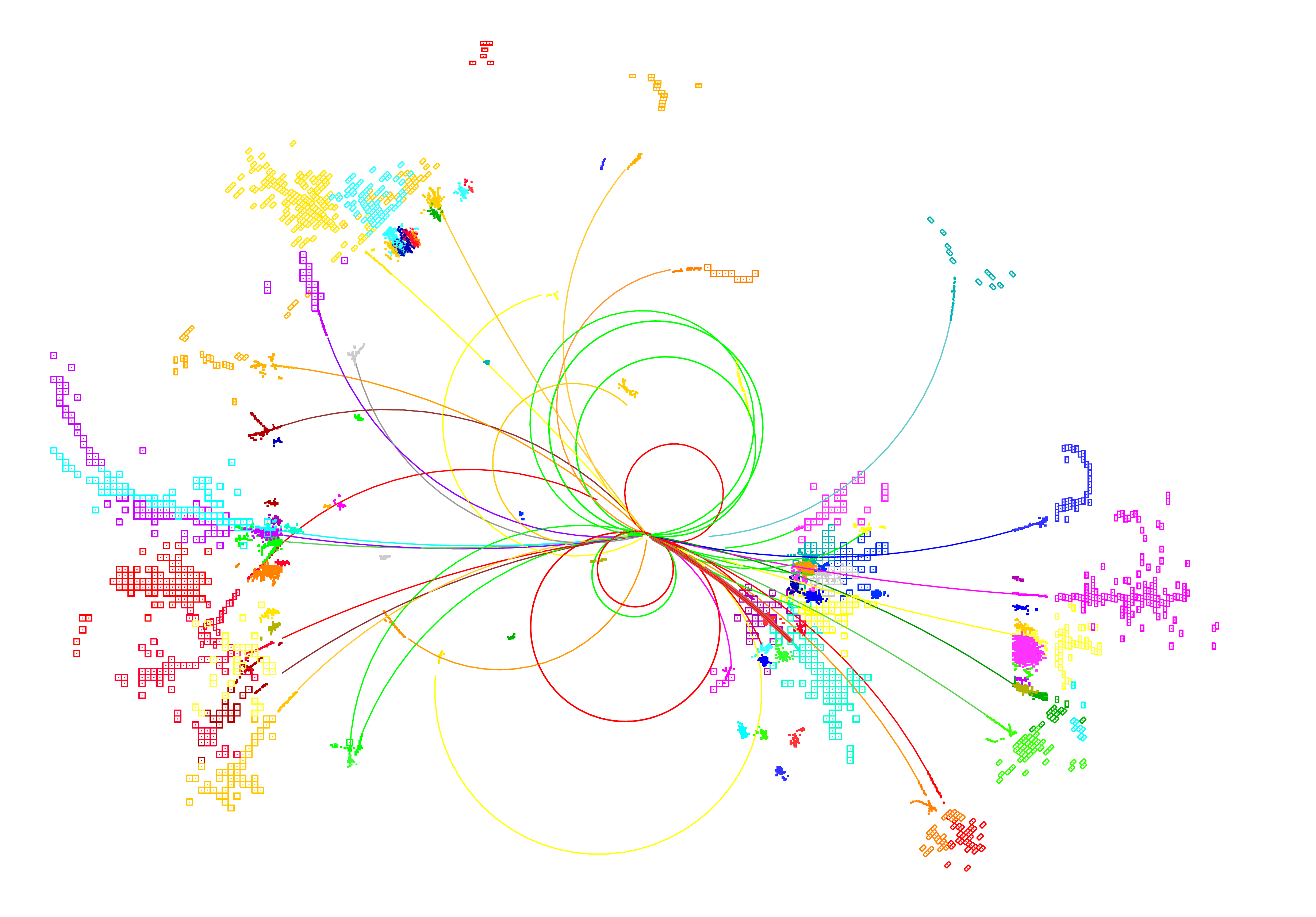}
     \label{fig::LCRecoAfter}}
     \caption{\label{fig::LCReco}Simulated detector response to a typical event, containing four jets, at CLIC. (a) The input Tracks and CaloHits and, (b), the output PFOs.}
  \end{center}
\end{figure}

The Pandora Linear Collider (LC) event reconstruction is performed by algorithms and plugins in the LCContent library. It aims to trace the paths of visible particles through a
fine granularity detector consisting of inner tracking detectors, an electromagnetic calorimeter (ECAL), a hadronic calorimeter (HCAL) and a series of muon chambers. A typical
event topology, containing multiple particles in a dense jet environment, is displayed in Figure \ref{fig::LCRecoBefore}.

The tracking detectors provide highly accurate space-point measurements. The pattern recognition and fitting of tracks in the tracking detectors is performed outside of Pandora, as described
in\,\cite{bib::TrackFitting}. Pandora Tracks represent the reconstructed trajectories and serve as the input to the particle flow algorithm. The ECAL and HCAL are sampling calorimeters, consisting of absorber
material, such as tungsten or steel, followed by layers of active material, such as silicon or scintillator, divided into individual cells. Pandora CaloHits represent energy deposits
in the calorimeter cells. Typical cell volumes are \mbox{$5\times5\times0.5\,\mathrm{mm}^{3}$} in the ECAL and \mbox{$3\times3\times0.3\,\mathrm{cm}^{3}$} in the HCAL. 

The LC event reconstruction is developed and tested using full GEANT4\,\cite{bib::GEANT4_1,bib::GEANT4_2} simulations of the ILD\,\cite{bib::ILC_TDR_4} and CLIC\_ILD\,\cite{bib::CLIC_ILD}
detector concepts in MOKKA\,\cite{bib::MOKKA}. Monte Carlo event samples for physics studies are generated using WHIZARD\,\cite{bib::WHIZARD}. Parton showering, hadronisation and fragmentation
is performed using PYTHIA\,\cite{bib::PYTHIA}. The Pandora LC client application is implemented in the MARLIN\,\cite{bib::MARLIN} software framework. Algorithm \ref{alg::MarlinPandora} describes
the operations that must be performed by this client application. 

\begin{algorithm}
  \begin{algorithmic}[1]
    \Procedure{main}{}
      \State Create a Pandora instance
      \State Register Algorithms and Plugins
      \State Provide detector geometry description
      \State Ask Pandora to parse XML settings file
      \ForAll{Events}
        \State Create Track instances
        \State Create CaloHit instances
        \State Create MCParticle instances
        \State Specify Track-Track relationships
        \State Specify MCParticle-Track relationships
        \State Specify MCParticle-CaloHit relationships
        \State Ask Pandora to process the event
        \State Get output PFOs and write to file
        \State Reset Pandora before next event
      \EndFor
    \EndProcedure
  \end{algorithmic}
  \caption{Pseudocode description of a client application used for Linear Collider event reconstruction.}\label{alg::MarlinPandora}
\end{algorithm}

The reconstruction proceeds as follows:

\begin{itemize}
   \item CaloHits are clustered using a simple cone-based algorithm, which works outwards from the interaction point, either adding CaloHits to existing Clusters or using them
   to create new Clusters. Clusters can be seeded by the projection of Tracks to the front face of the ECAL.
   \item The Clustering algorithm is configured so that it tends to split CaloHits from individual particles into multiple Clusters, rather than risk merging energy deposits
   from multiple particles into single Clusters. The Clusters are instead carefully merged together by a series of algorithms implementing well-defined topological rules.
   \item Clusters are associated to Tracks via careful comparison of Cluster positions and directions (obtained, for instance, via sliding linear fits) with projected Track
   positions and directions at the ECAL.
   \item The compatibility of associated Tracks and Clusters is assessed, via comparison of Track momentum with associated Cluster energies. Significant discrepancies indicate
   pattern recognition problems and the reclustering approach described in Section \ref{sec::Reclustering} is used to improve the clustering.
   \item Clusters without associated tracks are examined to assess whether they genuinely represent electrically neutral particles, or whether they are more likely to be fragments
   of any nearby track-associated Clusters, representing charged particles.
   \item PFOs are formed from Tracks and/or Clusters, representing the final pattern-recognition output. Particle identification plugins are used to label specific particle types.
   In the particle flow calorimetry approach, the properties of PFOs are extracted from Tracks when possible, otherwise they are determined from calorimeter information.
\end{itemize}

The results of processing a typical event are displayed in Figure \ref{fig::LCRecoAfter}.\\

The performance of the LC event reconstruction is typically characterised by measuring the jet energy resolution. This is determined using samples of Z' particles, which are
off-shell Z bosons, produced at rest at different centre of mass energies. The Z' particles decay into light quarks and produce two back-to-back mono-energetic jets. The resolution
of the jet energy, $E_{j}$, can be determined from the total reconstructed energy distribution, $E_{jj}$ as follows:

\begin{equation}
  \frac{\mathrm{RMS}_{90}(E_j)}{\mathrm{mean}_{90}(E_{j})} = \frac{\mathrm{RMS}_{90}(E_{jj})}{\mathrm{mean}_{90}(E_{jj})} \sqrt{2}
\end{equation}

\noindent where the RMS$_{90}$ is defined as the smallest RMS reconstructed in any region containing 90\,\% of the events. It is introduced in order to reduce sensitivity to tails in a
well-defined manner, because the effects of pattern recognition failures mean that the PFO energy distribution will be inherently non-Gaussian.

\begin{figure}[!h]
  \begin{center}
     \subfloat[][]{\includegraphics[width=0.45\textwidth]{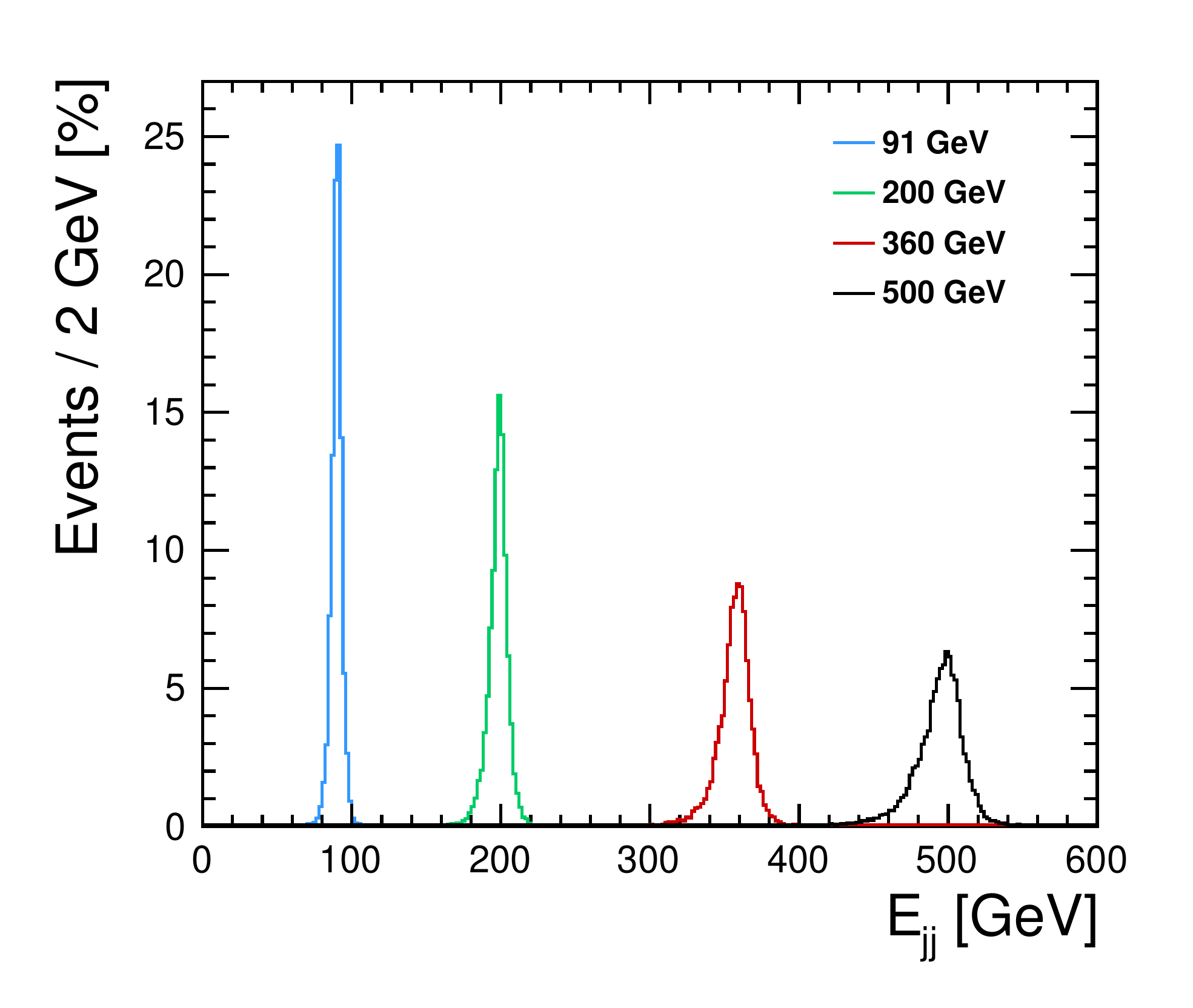}
     \label{fig::LCPerformance1}}\\
     \subfloat[][]{\includegraphics[width=0.45\textwidth]{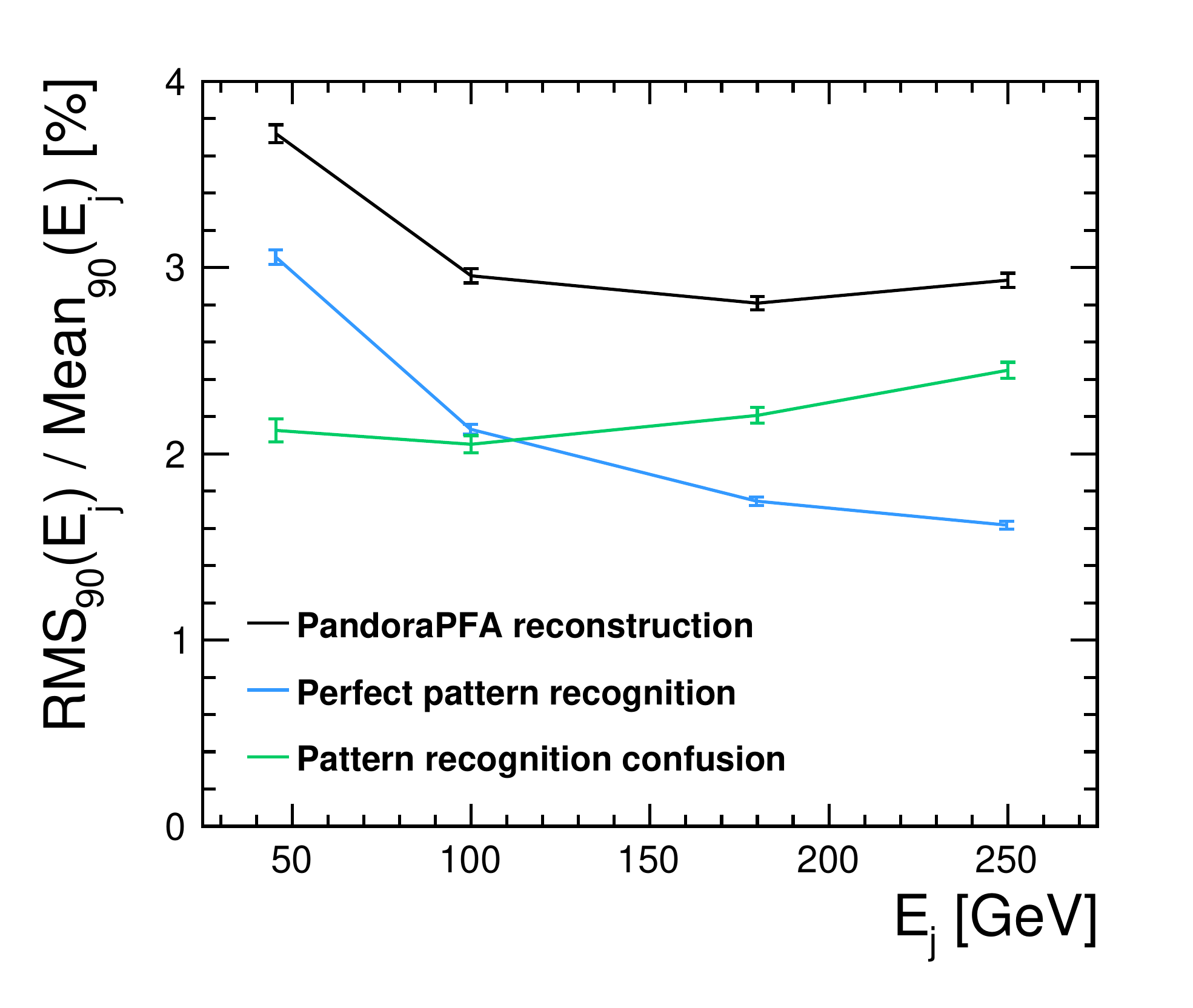}
     \label{fig::LCPerformance2}}
     \caption{\label{fig::LCPerformance}Performance of the Pandora LC event reconstruction for simulated Z' events in the ILD detector concept. (a) The total reconstructed energy
     for Z' events at different centre of mass energies. (b) The jet energy resolution as a function of jet energy, for jets in the barrel region of the detector. The contributions
     from the intrinsic calorimeter energy resolution and from pattern recognition ``confusion" are also indicated.}
  \end{center}
\end{figure}

Figure \ref{fig::LCPerformance1} shows the total reconstructed energy distributions for Z' events at different energies in ILD. Figure \ref{fig::LCPerformance2} shows the
variation of the jet energy resolution as a function of the jet energy, for jets in the barrel region of the detector. The same Figure shows the contributions to the jet energy
resolution from the intrinsic calorimeter energy resolution and from pattern recognition failures. The intrinsic energy resolution contribution is dominant at low jet energies,
but decreases with jet energy. The contribution from pattern recognition ``confusion" increases with jet energy and dominates at high jet energies. The jet energy
resolutions surpass the challenging ILC and CLIC targets of \mbox{$\sigma_{E}/E\,<\,3.5\textrm{--}5\,\%$}.

The CPU time and overall memory footprint for the LC event reconstruction is summarised in Table \ref{tab::LCPerformance}. A more detailed breakdown of the memory usage is shown in
Table \ref{tab::LCMemory}. %, which records details for configurations including a Pandora instance without any associated algorithm instances or input objects.
It can be seen that the Pandora instance itself has a rather small footprint, as do the algorithms. The input objects, for events with CLIC 3\,TeV backgrounds, are much more significant,
but it is the algorithm objects that prove most important. The optional ROOT-based monitoring functionality makes a significant contribution to the memory usage.

Table \ref{tab::LCCPU} demonstrates that the Pandora SDK operations represent a negligible amount of the LC event processing time. It is the operations performed within the algorithms
that dictate the event processing rate, as the algorithms collect evidence to inform their pattern recognition decisions. The most time-consuming SDK operation is the reset of
the Pandora instance between events.

\begin{table}[!h]
  \begin{center}
    \begin{tabular}{ l | c  c }
      \toprule
      Event type      & 200\,GeV Z'                 & 200\,GeV Z'                \\
      Background      & none                        & CLIC 3\,TeV                \\
      \midrule
      Event Time      & $0.22\pm0.02\,\mathrm{s}$   & $12.5\pm0.4\,\mathrm{s}$   \\
      VMem     	      & 256\,MB                     & 480\,MB                    \\
      RSS             & 43\,MB                      & 266\,MB                    \\
      \#\,Tracks      & $27\pm1$                    & $650\pm7$                  \\
      \#\,CaloHits    & $4,200\pm60$                & $39,900\pm700$             \\
      \bottomrule
    \end{tabular}
  \end{center}
  \caption{\label{tab::LCPerformance}Indicative CPU times and memory footprints for processing 200\,GeV Z' events in the CLIC detector, with and without
  overlaid backgrounds. The mean numbers of Tracks and CaloHits indicate the complexities of the events. The memory footprint is broken down into a virtual memory value and a resident
  set size value. The event times were recorded using a single socket, quad core, unthreaded Core i5-3570 CPU (clock speed 3.4GHz, SPECint2006 48.5, SpecFP206 62.9).}
\end{table}

\begin{table}[!h]
  \setlength{\tabcolsep}{4pt}
  \begin{center}
    \begin{tabular}{ c c c c c }
      \toprule
      Algorithms   & Input Objects & Monitoring   & VMem [MB] & RSS [MB]  \\
      \midrule
      \ding{55}    & \ding{55}     & \ding{55}    & 19        & 3         \\
      \ding{51}    & \ding{55}     & \ding{55}    & 19        & 4         \\
      \ding{55}    & \ding{51}     & \ding{55}    & 64        & 47        \\
      \ding{51}    & \ding{51}     & \ding{55}    & 244       & 229       \\
      \ding{51}    & \ding{51}     & \ding{51}    & 480       & 266       \\
      \bottomrule
    \end{tabular}
  \end{center}
  \caption{\label{tab::LCMemory}Memory footprint for processing 200\,GeV Z' events in the CLIC detector, with overlaid background. Virtual memory and resident set size values are
  shown for a number of configurations. These configurations indicate the separate memory footprints of the Pandora instance, the algorithm instances, the input objects and the
  algorithm objects (produced when both algorithms and input objects are provided). The memory footprint of the monitoring functionality is also indicated.}
\end{table}

\begin{table}[!h]
  \begin{center}
    \begin{tabular}{ l c c }
      \toprule
      Category      &  Operation Ranking(s) &  Processing Time [\%] \\
      \midrule
      Algorithm     &          1            &        12.0           \\
      Algorithm     &  1\,$\rightarrow$\,5  &        49.2           \\
      SDK           &          1            &         0.5           \\
      SDK           &  1\,$\rightarrow$\,5  &         0.9           \\
      \bottomrule
    \end{tabular}
  \end{center}
  \caption{\label{tab::LCCPU}Indication of the division of event processing time between algorithm and SDK operations for 200\,GeV Z' events in the CLIC detector, with overlaid background.
  The fraction of the time associated with the most time-consuming algorithm and SDK operations is shown. The cumulative fraction of time associated with the top-five most expensive
  algorithm and SDK operations is also displayed.}
\end{table}

\subsection{LAr TPC Event Reconstruction}
\begin{figure}[!h]
  \begin{center}
     \subfloat[][]{\includegraphics[width=0.45\textwidth]{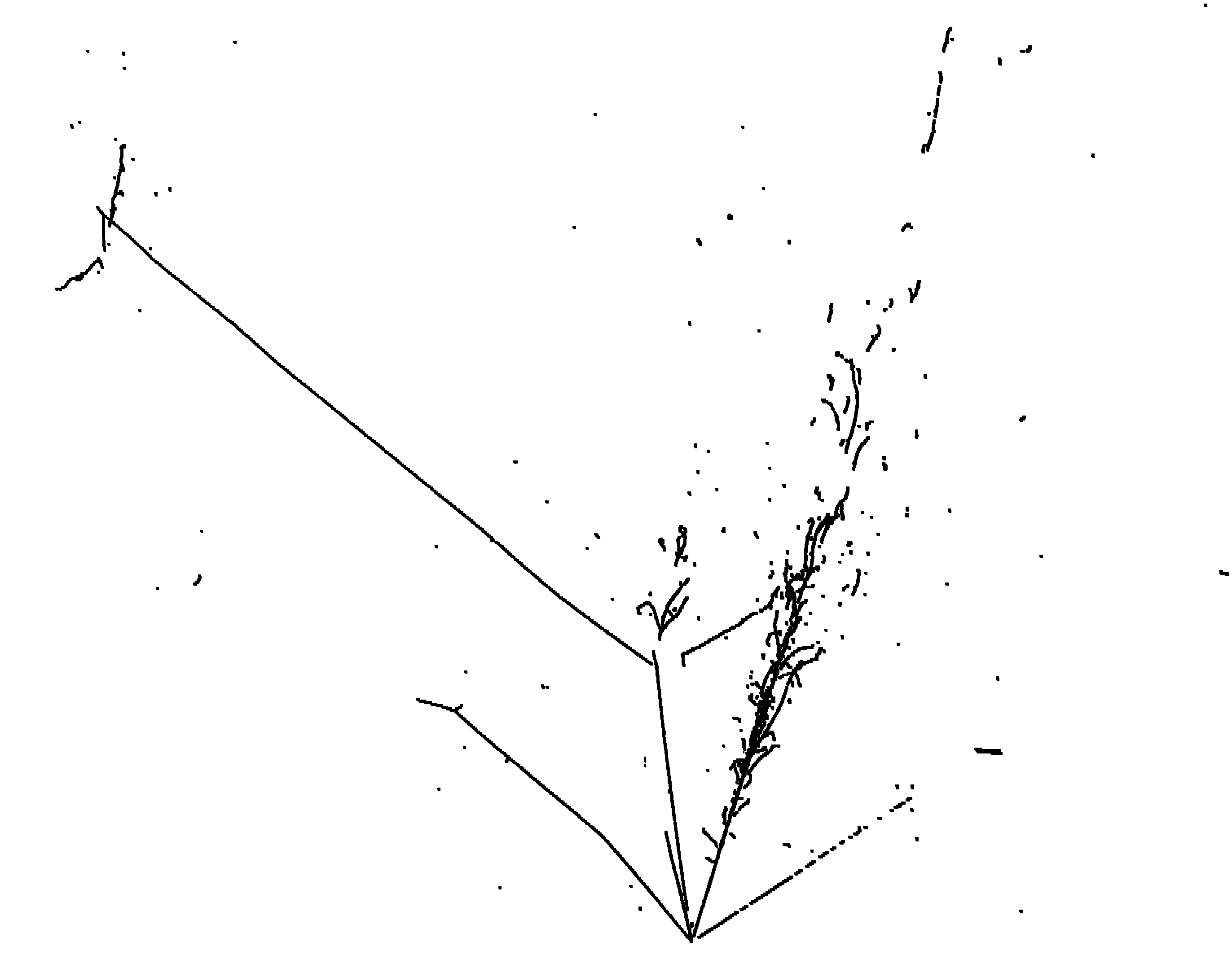}
     \label{fig::LArRecoBefore}}\\
     \subfloat[][]{\includegraphics[width=0.45\textwidth]{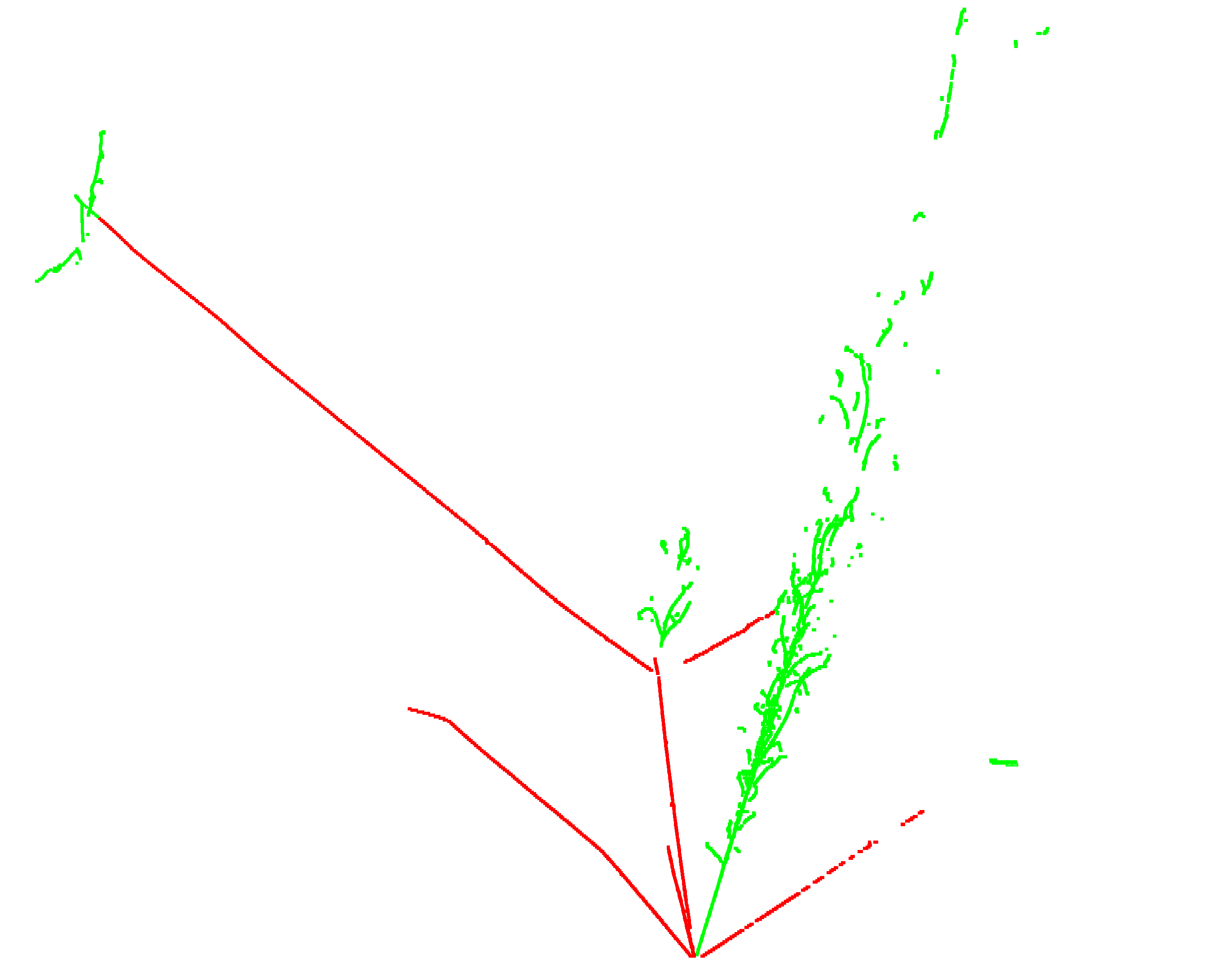}
     \label{fig::LArRecoAfter}}
     \caption{\label{fig::LArReco}Simulated detector response to a typical 5\,GeV CC $\nu_{e}$ event in a LAr TPC. (a) The input CaloHits and, (b), the output PFOs
     in one readout plane. Track-like PFOs, representing muons, protons or charged pions, are shown in red. Shower-like PFOs, representing electrons or photons,
     are shown in green.}
  \end{center}
\end{figure}

The Pandora LAr TPC event reconstruction is performed by algorithms, algorithm tools and plugins in the LArContent library. It aims to identify the paths of individual particles in
cosmic ray and neutrino-induced interactions in a LAr TPC with single-phase readout, where ionisation is detected using three readout (wire) planes in the liquid argon volume. The
readout provides three two dimensional images of the events within the detector active volumes. Each image shares a common coordinate, derived from the drift time (the difference
between the time at which ionisation signals were recorded and the ``$t_{0}$" identified for the event). The second coordinate is derived from the number of the wire recording the
ionisation signal. The readout pitch is of the order of mm, so each two dimensional image, such as that in Figure \ref{fig::LArRecoBefore}, provides a wealth of information.

The LAr TPC event reconstruction is developed using detailed GEANT4-based Monte Carlo simulations of the MicroBooNE\,\cite{bib::MicroBooNE} detector and the DUNE\,\cite{bib::DUNE} far detector
and 35-ton prototype. Events are generated using the GENIE\,\cite{bib::GENIE} simulation of neutrino-nucleus interactions and the CRY\,\cite{bib::CRY} cosmic ray generator. The Pandora LAr TPC
client application is implemented in LArSoft\,\cite{bib::LArSoft}. The reconstruction proceeds as follows:

\begin{itemize}
   \item The client application creates separate CaloHits for each of the recorded hits in the three images. The CaloHits can be distinguished via their HitType member variables
   and are all placed in the $y\,=\,0$ plane. The first algorithm filters the CaloHits into three separate lists based on their HitTypes.
   \item Reconstruction of two dimensional Clusters is performed separately for each of the three CaloHit lists. The two dimensional reconstruction begins with a track-based
   clustering algorithm, which searches for continuous, unambiguous lines of CaloHits. Separate Clusters are formed for each ``branch" visible in the CaloHit topologies.
   \item The Clustering algorithm is careful to avoid merging energy deposits from multiple particles into single Clusters. The two dimensional Clusters are instead carefully
   merged together via a series of algorithms looking for clear evidence of topological association between pairs or chains of Clusters.
   \item The three lists of two dimensional Clusters are used to reconstruct the most likely three dimensional event vertex position. Comparison of Clusters in each of the
   different readout planes is possible via a Pandora plugin that performs all required coordinate transformations.
   \item Reconstruction of three dimensional tracks is performed by examining the compatibility of combinations of two dimensional Clusters from each of the three readout
   planes. A rank-three tensor is created to store details of the suitability of all possible Cluster combinations. This tensor is interrogated by a set of algorithm tools,
   which identify ambiguities and make changes to the two dimensional reconstruction until the Cluster combinations are unambiguous (the tensor is diagonal).
   \item Consistent groups of the two dimensional Clusters are stored in PFOs, which provide a convenient means for collecting together objects reconstructed in the three
   separate readout planes.
   \item Shower reconstruction is performed. This begins in two dimensions, looking for associations between long ``shower-spine" Clusters and short ``shower-branch" Clusters.
   Shower Clusters from each of the three readout planes are compared and grouped-together in three dimensions using a further tensor-diagonalisation approach.
   \item Vertices are created for Track and Shower PFOs. Particle parent-daughter links are used in order to provide a reconstructed decay hierarchy. New three dimensional
   CaloHits and Clusters are created for each PFO as the ultimate reconstructed representation of the underlying event.
\end{itemize}

The results of processing a typical event are displayed in Figure \ref{fig::LArRecoAfter}\\

\begin{figure}[!h]
  \begin{center}
     \subfloat[][]{\includegraphics[width=0.45\textwidth]{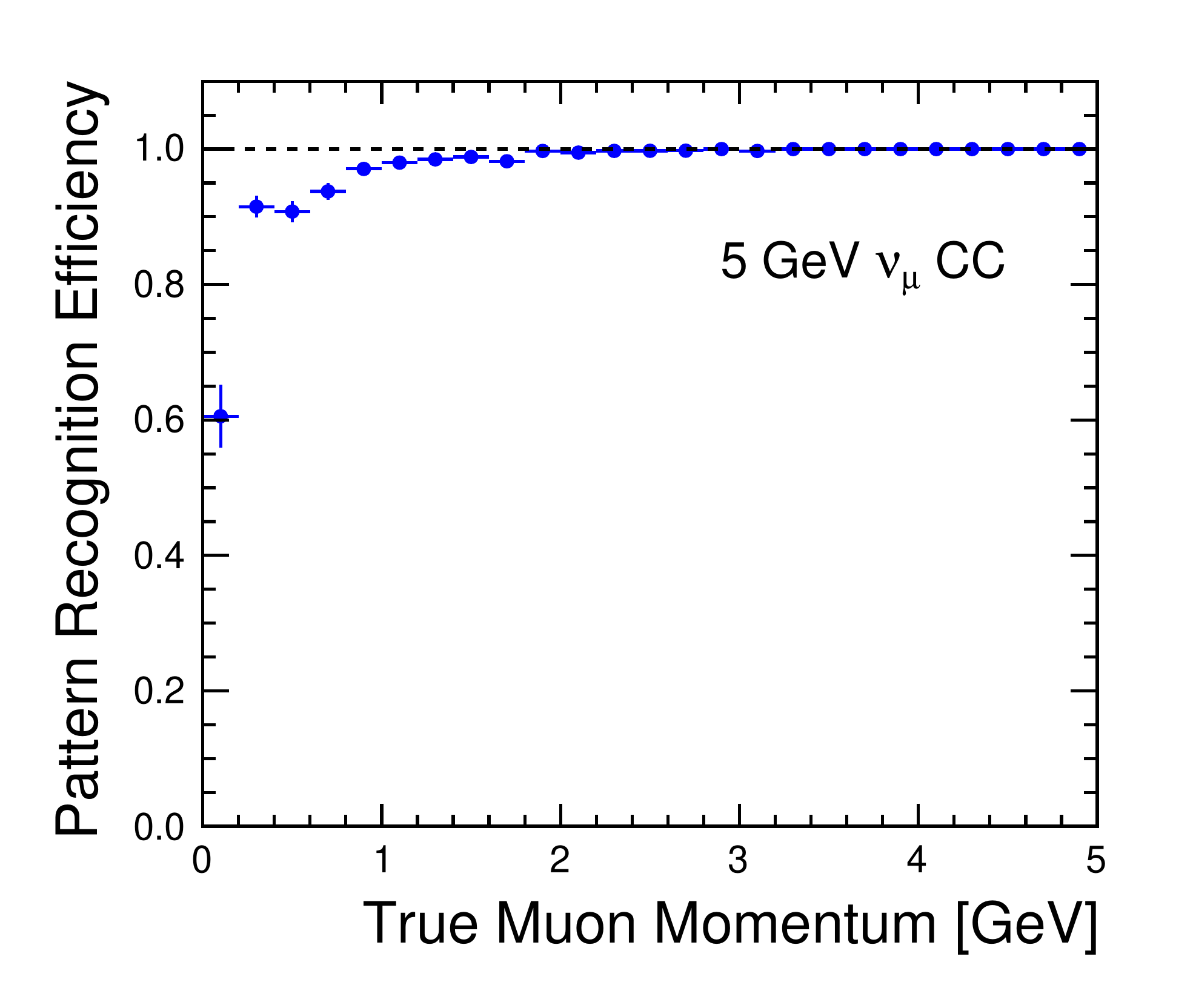}
     \label{fig::LArPerformance1}}\\
     \subfloat[][]{\includegraphics[width=0.45\textwidth]{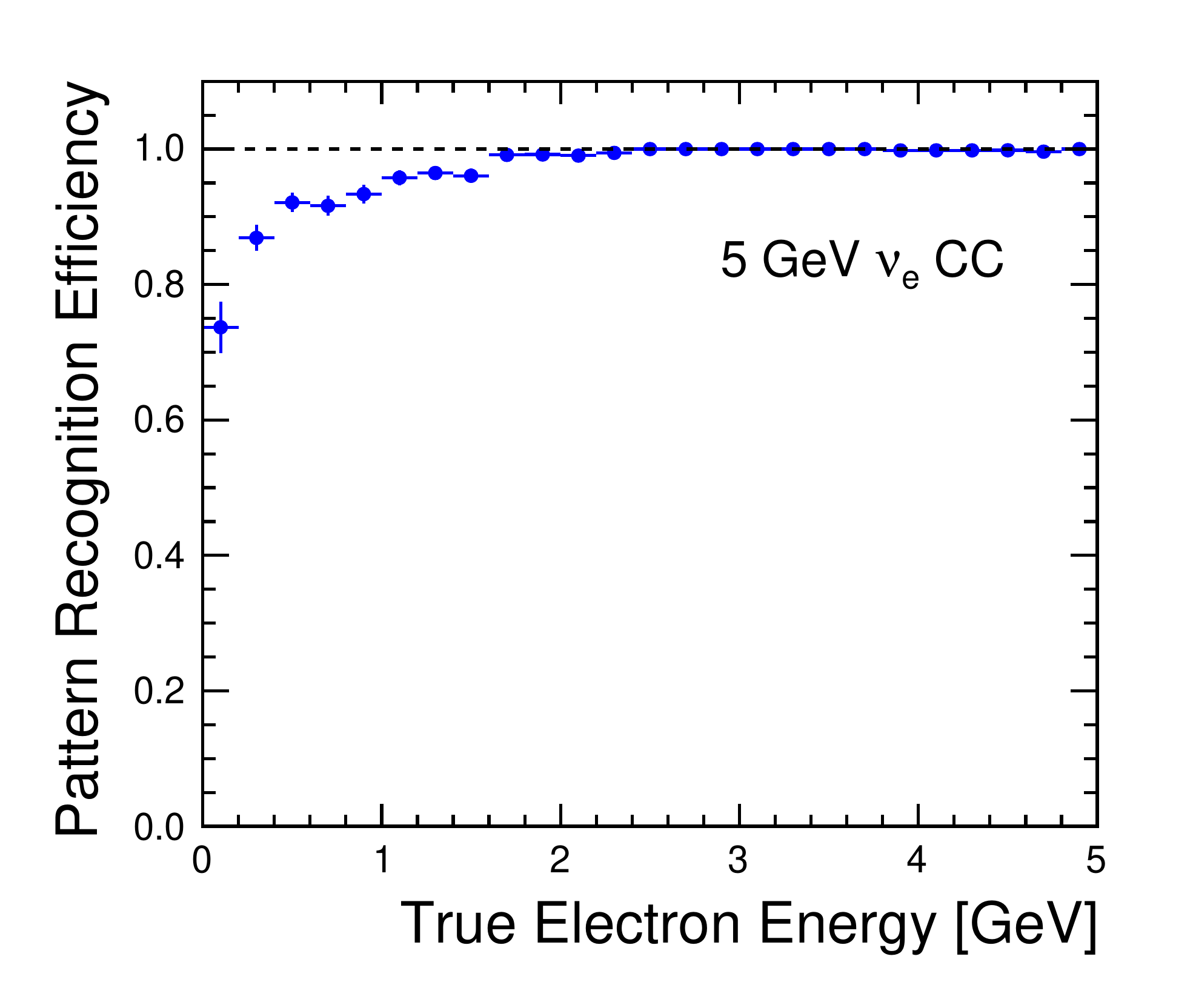}
     \label{fig::LArPerformance2}}
     \caption{\label{fig::LArPerformance}The reconstruction efficiency of the Pandora LAr TPC algorithms for leading final-state leptons in, (a), 5\,GeV $\nu_{\mu}$ CC and, (b),
     5\,GeV $\nu_{e}$ CC interactions simulated in the MicroBooNE detector.}
  \end{center}
\end{figure}

The performance of the LAr TPC reconstruction can be assessed via the efficiency for reconstructing final-state particles in the simulated events. Figure \ref{fig::LArPerformance}
shows the reconstruction efficiency for the leading lepton %, as a function of its momentum or energy, 
in 5\,GeV $\nu_{\mu}$ and $\nu_{e}$ charged current (CC) interactions in the MicroBooNE
detector. Successful reconstruction requires accurate clustering in all three readout planes and successful matching of clusters between planes. The efficiency is greater than 95\% for 
lepton momenta above 1\,GeV and approaches 100\% above 2\,GeV. The CPU time and memory footprint for the LAr TPC event reconstruction is summarised in Table \ref{tab::LArPerformance}.

\begin{table}[!h]
  \begin{center}
    \begin{tabular}{ l | c  c }
      \toprule
      Event type        & 5\,GeV $\nu_{e}$ CC        & BNB spectrum               \\
      Background        & none                       & cosmic ray                 \\
      \midrule
      Event Time        & $1.82\pm0.07\,\mathrm{s}$  & $3.18\pm0.15\,\mathrm{s}$  \\
      VMem              & 273\,MB                    & 446\,MB                    \\
      RSS               & 59\,MB                     & 232\,MB                    \\
      \#\,CaloHits      & $6,100\pm80$               & $14,600\pm500$             \\
      \bottomrule
    \end{tabular}
  \end{center}
  \caption{\label{tab::LArPerformance}Indicative CPU times and memory footprints for processing two different classes of events, simulated in the MicroBooNE detector. Note that the
  algorithm configuration differs for processing events with and without cosmic ray overlay. The event times were recorded using a single socket, quad core, unthreaded Core i5-3570
  CPU (clock speed 3.4GHz, SPECint2006 48.5, SpecFP206 62.9).}
\end{table}

%%%%%%%%%%%%%%%%%%%%%%%%%%%%%%%%%%%%%%%%%

\section{Concluding Comments}
The Pandora SDK provides an efficient and reusable software solution for developing pattern recognition algorithms. The SDK was developed to address problems in the field of High
Energy Physics. The Pandora LCContent library provides the state of the art in particle flow analysis for events in fine granularity detectors, such as those designed for
use at the proposed high-energy $e^{+}e^{-}$ linear colliders ILC and CLIC. The LC algorithms are also helping to drive detector optimisation studies for the upgrade of the CMS detector
at the LHC. In the neutrino sector, the Pandora LArContent library provides an advanced reconstruction of cosmic ray and neutrino-induced events in LAr TPCs and is used by the
MicroBooNE and DUNE experiments.

\section*{Acknowledgements}
This work was funded in part by the UK Science and Technology Facilities Council and by the European Union under the Advanced European Infrastructures for Detectors and 
Accelerators (AIDA) project. The authors would like to acknowledge the help provided by Peter Speckmayer, aiding the development of the Pandora visualisation, and Andrew Blake,
aiding the development of algorithms for use with events in LAr TPCs.

\end{document}